\documentclass[aps,prb,showpacs,superscriptaddress,twocolumn,amsmath,amssymb]{revtex4}
\usepackage{graphicx}
\usepackage{dcolumn}
\usepackage{multirow}
\usepackage{diagbox}

\usepackage{amsmath}
\usepackage{subfigure}
\usepackage{color}
\usepackage{epstopdf}
\usepackage[normalem]{ulem}

\usepackage{etoolbox}
\usepackage{etoolbox}
\usepackage[capitalize]{cleveref}
\usepackage{graphicx}
\usepackage{float}
\usepackage{dcolumn}
\usepackage{lipsum}
\usepackage{bm}
\usepackage{epsfig}
\usepackage{epstopdf}
\usepackage{blindtext}
\usepackage{subfigure}
\usepackage{dcolumn}
\usepackage{ifthen}
\usepackage{threeparttable}
\usepackage{longtable}
\include{newcommands}

\begin{document} 
\title{Comparing particle-particle and particle-hole channels of random-phase approximation}

\author{Muhammad N. Tahir}
\affiliation{CAS Key Laboratory of Quantum Information, University of Science and Technology of China, Hefei, Anhui 230026, China}
\affiliation{CAS Center For Excellence in Quantum Information and Quantum Physics, University of Science and Technology of China, Hefei, Anhui 230026, China}

\author{Xinguo Ren}
\affiliation{CAS Key Laboratory of Quantum Information, University of Science and Technology of China, Hefei, Anhui 230026, China}
\affiliation{CAS Center For Excellence in Quantum Information and Quantum Physics, University of Science and Technology of China, Hefei, Anhui 230026, China}

\date{\today }

\begin{abstract}
We present a comparative study of particle-hole and particle-particle channels of random-phase approximation (RPA) for molecular dissociations
of different bonding types. We introduced a \textit{direct} particle-particle RPA scheme, in analogy to the \textit{direct} particle-hole RPA formalism, 
whereby the exchange-type contributions are excluded. This allows us to compare the behavior of the particle-hole and particle-particle RPA channels
on the same footing. Our study unravels the critical role of exchange contributions in determining behaviors of the two RPA channels for
describing stretched molecules. We also made an attempt to merge particle-hole RPA and particle-particle RPA into a unified scheme, with the double-counting
	terms removed. However, benchmark calculations indicate that a straightforward combination of the two RPA channels does not lead to a 
	successful computational scheme for describing molecular dissociations.
\end{abstract}

\maketitle

\section{Introduction}
In the past two decades, the random-phase approximation (RPA), originally formulated by Bohm and Pines \cite{Bohm/Pines:1951,Bohm/Pines:1953} for
the homogeneous gas of interacting electrons, has been developed
into a versatile approach to compute the non-local electron correlation energy in real
molecules and materials \cite{Hesselmann/Goerling:2011,Eshuis/Bates/Furche:2012,Ren/etal:2012b,Paier/etal:2012}. Successful applications of RPA
have been demonstrated for molecules \cite{Furche:2001,Fuchs/Gonze:2002,Toulouse/etal:2009,Zhu/etal:2010,Hesselmann/Goerling:2011b,Ren/etal:2011,Eshuis/Furche:2011}, 
solids \cite{Harl/Kresse:2008,Harl/Kresse:2009,Harl/Schimka/Kresse:2010,Lebegue/etal:2010,Nguyen/deGironcoli:2009,Lu/Li/Rocca/Galli:2009,Casadei/etal:2012,Casadei/etal:2016}, surfaces \cite{Ren/etal:2009,Schimka/etal:2010}, interfaces \cite{Mittendorfer/etal:2011,Olsen/etal:2011}, 
and defects \cite{Bruneval:2012,Kaltak/Klimes/Kresse:2014,Kaltak/Klimes/Kresse:2014b}. 
As such, RPA sets a new stage for first-principles electronic-structure calculations for real materials. From the methodology point of view,
RPA represents a cornerstone connecting ground-state and time-dependent density-functional theory 
(DFT)\cite{Langreth/Perdew:1975,Gunnarsson/Lundqvist:1976,Dobson:1994}, 
the quantum chemistry coupled cluster method \cite{Scuseria/Henderson/Sorensen:2008}, and the Green-function based 
many-body perturbation theory \cite{Dahlen/Leeuwen/Barth:2005,Ren/etal:2012b}. 
In the context of DFT, via the adiabatic connection fluctuation dissipation theorem, the RPA formalism opens an arena to construct advanced
exchange-correlation functional in terms of diagrammatic many-body perturbation theory.  Further development
beyond RPA, aiming at addressing its remaining shortcomings, has been an active research area. The renormalized second-order perturbation theory 
\cite{Paier/etal:2012,Ren/etal:2013}, the approximate exchange kernel (AXK) correction \cite{Bates/Furche:2013,Guo/Agee/Furche:2018},  and 
the power-series expansion scheme \cite{Erhard/Bleiziffer/Goerling:2016} represent the latest developments based on RPA.

In nuclear physics, the above-mentioned RPA approach is referred to as particle-hole RPA (phRPA), because in its formulation the correlation energy is determined 
in terms of density fluctuations, originating from particle-hole (ph) pair excitations. In parallel to the phRPA, another type of RPA formulation, referred to as
particle-particle RPA (ppRPA), is also discussed \cite{Ring/Schuck:1980, Blaizot/Ripka:1986}. In this case, the correlation energy 
is obtained from the pairing matrix fluctuation, arising from the process of creating two particles or two holes. From a diagrammatic point of view, 
phRPA can be viewed as a summation of ring diagrams to infinite order \cite{Gell-Mann/Brueckner:1957}, whereas the ppRPA can be interpreted as a summation of
the so-called ladder diagrams to infinite order \cite{Ring/Schuck:1980,Blaizot/Ripka:1986}. 
In a series of seminal papers, Yang and coworkers \cite{Aggelen/etal:2013,Peng/Steinmann/etal:2013,Yang/Aggelen/Steinmann/etal:2013} developed an adiabatic 
connection formalism which allows to express the exchange-correlation energy in terms of dynamical pairing matrix fluctuation \cite{Aggelen/etal:2013}. 
Within such a formulation, the ppRPA is the leading-order approximation. The performance of ppRPA for thermochemistry has been 
benchmarked for small molecules, and its quality for describing molecular dissociations has been analyzed 
in terms of fractional charge and fractional spin errors \cite{Yang/Aggelen/Steinmann/etal:2013,Aggelen/etal:2014}. Furthermore, in parallel to time-dependent
DFT (TDDFT), the ppRPA formulation has been extended to {excited} state calculations \cite{Peng/etal:2014}. Through these pioneering works, Yang and coworkers
brought ppRPA to the attention of DFT/materials science communities.  The equivalence of ppRPA to ladder coupled cluster doubles theory have been 
demonstrated independently by Yang's Group \cite{Peng/Steinmann/etal:2013} and Scuseria's group \cite{Scuseria/Henderson/Bulik:2013}.

Given the fact that there exists two RPA channels, it is natural to ask if it is possible to combine them. From the viewpoint of many-body perturbation theory,
they contain distinct diagrams, except at the second order.  In a sense, phRPA and ppRPA can be viewed as different ways to ``renormalize" 
the bare second-order correlation energy. For correlated methods, the correlation energy can often be linked to a
self-energy (obtained by taking the functional derivative of the correlation energy with respect to the Green function) \cite{Luttinger/Ward:1961}. For instance, 
the phRPA correlation energy is intimately connected to the $GW$ self-energy \cite{Hedin:1965,Ren/etal:2012b}; the ppRPA, on the other hand, is associated with the
particle-particle (pp) T-matrix approximation \cite{Baym/Kadanoff:1961,Kanamori:1963,Romaniello/Bechstedt/Reining:2012,Zhang/Su/Yang:2017}.
Physically speaking, phRPA accounts for the non-local screening effect arising from the long-range Coulomb interaction and shows best performance in the 
high electron density regime.  In contrast, the ppRPA (and T-matrix approximation) describes local scattering of hard-core potentials \cite{Bethe/Goldstone:1957}, 
and represents a good approximation in the low electron density regime. 
In the past, attempts have been made to combine the $GW$ approximation and T-matrix approximation for self-energy 
calculations \cite{Romaniello/Bechstedt/Reining:2012}.  It is interesting to check if it is possible to merge 
their counterparts for ground-state correlation energies -- phRPA and ppRPA -- into one framework. In this work, we implemented ppRPA in the FHI-aims code 
package \cite{Blum/etal:2009,Ren/etal:2012}. Together with our earlier implementation of the phRPA, we are able to make an attempt to check 
if a useful computational framework can be found by combining phRPA and ppRPA, and moreover, compare the behavior of phRPA and ppRPA in a systematic way. 

When comparing the performance of phRPA and ppRPA for molecules' properties, it is important to note that 
phRPA by default refers to the \textit{direct} phRPA, without including the exchange contributions, whereas ppRPA refers to
the \textit{full} ppRPA, with the exchange contributions included. To compare phRPA and ppRPA on the same footing, we also implemented in this work
the \textit{full} phRPA in FHI-aims. This allows us to examine and compare the performance of phRPA and ppRPA for 
molecular dissociations in an unbiased way.

The rest of the paper is organized as follows. We briefly recapitulate the basic equations of phRPA and ppRPA in Sec.~\ref{sec:theory}. This is followed by
a description of the implementation and computational details in Sec.~\ref{sec:comput}. The major results and discussions of the behavior of the two channels
of RPA, as well as their combinations, are presented in Sec.~\ref{sec:results} for prototypical molecular dimers. Finally{,} we conclude this work in 
Sec.~\ref{sec:conclusion}.

\section{\label{sec:theory}Theory}
Here we briefly review the basic equations of phRPA and ppRPA to facilitate the subsequent comparative analysis of the two schemes.
More detailed accounts on their theoretical foundations can be found, e.g., in Refs.~[\onlinecite{Furche:2001,Fuchs/Gonze:2002,Hesselmann/Goerling:2011,Eshuis/Bates/Furche:2012,Ren/etal:2012b,Aggelen/etal:2013}].  Note that both the phRPA and the ppRPA can be formulated as an approximation to the exchange-correlation energy 
within Kohn-Sham (KS) DFT via the adiabatic connection framework.
While the phRPA correlation energy can be expressed in terms of an integration over the polarization propagator, the ppRPA can be analogously obtained from
the pp propagator, or equivalently the pairing density fluctuation \cite{Aggelen/etal:2013}. In particular, the ppRPA correlation
energy is given by,
\begin{equation}
	E_{c}^\text{ppRPA}=\dfrac{1}{2\pi i}  \int^{+\textit{i}\infty} _{-\textit{i}\infty} 
	\text{Tr}[ln[\textbf{I}-\textbf{K}^{0}(\omega)\textbf{V}] + \textbf{K}^{0}(\omega) 
\textbf{V}]d\omega \, ,
\label{Eq:ppRPA_Ec}
\end{equation}
where $\textbf{K}^{0}(\omega)$ is the non-interacting pp propagator,
 \begin{equation}
 \begin{split}
 \textbf{K}^{0}_{pq,rs}(\omega)=& \quad  \left(\delta_{pr}\delta_{qs} -\delta_{ps}\delta_{qr}\right) \dfrac{\theta(\epsilon_{p}-\epsilon_F)\theta(\epsilon_{q}-\epsilon_F)}{\omega-(\epsilon_{p}+\epsilon_{q}-2\mu)+\textit{i}\eta}\\&-
 \left(\delta_{pr}\delta_{qs}  -\delta_{ps}\delta_{qr}\right) \dfrac{\theta(\epsilon_{F}-\epsilon_{p})\theta(\epsilon_{F}-\epsilon_{q})}{\omega-(\epsilon_{p}+\epsilon_{q}-2\mu)-\textit{i}\eta} \, .
 \end{split}
 \end{equation}
with $\theta(x)$ and $\delta_{pr}$ being respectively the Heaviside step function and Kronecker Delta function. Furthermore, $\epsilon_{F}$ is the Fermi energy, 
and $\mu$ is the chemical potential set at the middle of the gap between the highest occupied molecular orbital (HOMO) and lowest unoccupied molecular orbital (LUMO). 
Moreover, $\textbf{V}$ is the antisymmetrized 
two-electron Coulomb integrals, 
 \begin{equation*}
\textbf{V}_{pq,rs}=\langle pq||rs\rangle=\langle pq|rs\rangle-\langle pq|sr\rangle 
 \end{equation*}
and
\begin{equation} \label{columb_integral}
\langle pq|rs \rangle= \int \dfrac{\phi^{\ast} _{p} (x_{1}) \phi^{\ast}_{q}(x_{2})\phi_{r}(x_{1})\phi_{s}(x_{2})}{|\bf{r}_{1}-\bf{r}_{2}|} dx_{1}dx_{2}
\end{equation} 
with $x_1=(\bf{r}_1,\sigma_1)$ being the combined spatial-spin coordinate. Here we follow the usual convention that $p,q,r,s$ refer to general single-particle
spin-orbitals, whereas $i,j,k,l$ and $a,b,c,d$ being occupied and unoccupied spin-orbitals respectively.

As thoroughly discussed in the literature \cite{Furche:2001,Eshuis/Bates/Furche:2012,Ren/etal:2012b}, the phRPA correlation energy has a similar expression
as Eq.~(\ref{Eq:ppRPA_Ec}). The key difference is that the non-interacting pp propagator $\textbf{K}^{0}(\omega)$ is replaced by the
non-interacting polarization operator (or linear density response function) ${\chi}^0$,
\begin{equation}
 \begin{split}
 {\chi}^{0}_{pq,rs}(\omega)=& \quad  \delta_{pr}\delta_{qs} \dfrac{\theta(\epsilon_{p}-\epsilon_F)\theta(\epsilon_F-\epsilon_{q})}{\omega-(\epsilon_{p}-\epsilon_{q})+\textit{i}\eta}\\&-
 \delta_{pr}\delta_{qs} \dfrac{\theta(\epsilon_{F}-\epsilon_{p})\theta(\epsilon_{q}-\epsilon_{F})}{\omega-(\epsilon_{p}-\epsilon_{q})-\textit{i}\eta} \, .
 \end{split}
\end{equation}
Furthermore, if the Coulomb matrix $\textbf{V}$ is not antisymmetrized (i.e., $\textbf{V}_{pq,rs}=\langle pq|rs\rangle$), 
one will obtain the so-called \textit{direct} phRPA ($d$-phRPA),
which is the standard RPA method employed in density functional/materials science community. In contrast, if the antisymmetrized $\textbf{V}$ 
is employed, one will have the  \textit{full} phRPA ($f$-phRPA), which however received much less attention and was only discussed in the quantum 
chemistry literature \cite{McLachlan/Ball:1964,Oddershede/etal:1975,Angyan/etal:2011,Rolf/Seeger:1977}. 
Different from the phRPA, the ppRPA are only implemented and discussed with the antisymmetrized Coulomb interaction \cite{Aggelen/etal:2013,Peng/Steinmann/etal:2013,Yang/Aggelen/Steinmann/etal:2013,Aggelen/etal:2014}. In analogy to phRPA, in this work we shall term
the standard ppRPA with antisymmetrized Coulomb integrals the \textit{full} ppRPA ($f$-ppRPA), and the one without antisymmetrizing the Coulomb integrals as 
\textit{direct} ppRPA ($d$-ppRPA). As mentioned above, this enables us to benchmark phRPA and ppRPA on an equal footing, separating the effect arising
from the exchange interactions and that arising from the ph or pp channel itself.

The RPA correlation energy can be calculated using  Eq.~(\ref{Eq:ppRPA_Ec}), with the frequency integration being carried out
along the imaginary axis. In case of the \textit{direct} phRPA, the imaginary frequency integration combined with the resolution-of-identity approximation
\cite{Eshuis/Yarkony/Furche:2010,Ren/etal:2012} leads to an efficient $O(N^4)$ scaling algorithm, with $N$ being the basis size of the system. An alternative way 
to obtain the phRPA correlation energy is to cast the RPA equations into the following generalized eigenvalue problem 
\cite{McLachlan/Ball:1964,Furche:2001,Scuseria/Henderson/Sorensen:2008,Mori-Sanchez/etal:2012}, 
\begin{equation}
\begin{pmatrix}
\textbf{A} & \textbf{B} \\ 
 \textbf{B$^{\ast}$} & \textbf{A$^{\ast}$}
\end{pmatrix} 
\begin{pmatrix}
X \\ Y
\end{pmatrix} = \omega_n
\begin{pmatrix}
\textbf{1} & 0 \\
0 & -\textbf{1}
\end{pmatrix} 
\begin{pmatrix}
X \\ Y
\end{pmatrix}
\label{eq:phRPA_eigen}
\end{equation}
where $A$, $B$, $X$, $Y$ are square matrices of dimension $N_h N_p\times N_h N_p$, with $N_h$ and $N_p$ being
the number of occupied (hole) and unoccupied (particle) orbitals respectively. The eigenvalues $\omega_n$ form
a vector of dimension $N_h N_p$, and correspond to the neutral excitation energies at the RPA level. 
In case of \textit{full} phRPA, 
\begin{align}
	A_{ia,jb} & =\langle \Phi_i^{a}|\hat{H} - E_0| \Phi_j^b \rangle =\delta_{ij}\delta_{ab} (\epsilon_{a}-\epsilon_{i})+\langle aj||ib \rangle \nonumber \\
	B_{ia,jb} & = \langle \Phi_0|\hat{H} - E_0| \Phi_{ij}^{ab} \rangle = \langle ij|| ab \rangle 
	\label{f-phrpa_AB}
\end{align}
where $\hat{H}$ is the Hamiltonian of interacting electrons, $E_0 = \langle \Phi_0 | \hat{H} | \Phi_0 \rangle $ is the Hartree-Fock ground-state energy,
with $\Phi_0$ being the lowest-energy single Slater determinant. In Eq.~(\ref{f-phrpa_AB}), $\Phi_i^a$ and $\Phi_{ij}^{ab}$ are singly and doubly excited configurations, respectively.
The corresponding \textit{full} phRPA (\textit{f}-phRPA) correlation energy can be obtained as \cite{McLachlan/Ball:1964}
\begin{equation}
E^{\textit{f}\textnormal{-phRPA}}_c =\frac{1}{4}\left( \sum_{n}\omega_n - Tr\{A\} \right) \,.
\label{eq:Ec_fphRPA}
\end{equation}
By contrast, the \textit{direct} phRPA excitation energies and amplitudes can be obtained by solving Eq.~(\ref{eq:phRPA_eigen}) with 
\begin{align}
 A_{ia,jb} & =\delta_{ij}\delta_{ab} (\epsilon_{a}-\epsilon_{i})+\langle aj|ib \rangle  \nonumber \\
 B_{ia,jb} & =\langle ij| ab \rangle
\end{align}
i.e., without antisymmetrizing the two-electron Coulomb integrals in the construction of $A$ and $B$. Now the $d$-phRPA correlation energy is given by
\begin{equation}
E^{\textit{d}\textnormal{-phRPA}}_c =\frac{1}{2}\left( \sum_{n}\omega_n - Tr\{A\} \right) \, .
\label{eq:Ec_dphRPA}
\end{equation}
Note that the choice of different prefactors in Eq.~(\ref{eq:Ec_fphRPA}) and (\ref{eq:Ec_dphRPA}) is to ensure that both theories 
have the correct behavior at second order \cite{McLachlan/Ball:1964,Angyan/etal:2011,Hesselmann:2011}. 
However, the choice of a factor of $1/2$ in Eq.~(\ref{eq:Ec_fphRPA}) has also been used in the literature \cite{Scuseria/Henderson/Bulik:2013}. 
The meaning of $d$-phRPA correlation energy can be interpreted as the difference of correlated and uncorrelated electronic zero-point plasmonic
energies \cite{Furche:2008}.

The ppRPA, instead, can be cast into the following matrix equation \cite{Aggelen/etal:2013,Scuseria/Henderson/Bulik:2013},
\begin{equation}
\begin{pmatrix}
\textbf{C} & \textbf{B} \\ 
 \textbf{B}^{\dagger} & \textbf{D}
\end{pmatrix} 
\begin{pmatrix}
X \\ Y
\end{pmatrix} = \omega_n
\begin{pmatrix}
\textbf{1} & 0 \\
0 & -\textbf{1}
\end{pmatrix} 
\begin{pmatrix}
X \\ Y
\end{pmatrix}
\label{Eq:ppRPA_matrix}
\end{equation}
where 
\begin{align}
C_{ab,cd}&=\langle \Psi_0 | \hat{c}_a \hat{c}_b \left(\hat{H} - E_0 \right) \hat{c}^{\dagger}_c \hat{c}^{\dagger}_d |\Phi_0 \rangle \nonumber \\
           & = \delta_{ac}\delta_{bd} (\epsilon_{a}+\epsilon_{b}-2\mu)+\langle ab||cd \rangle \\
D_{ij, kl} &=\langle \Psi_0 | \hat{c}^{\dagger}_i \hat{c}^{\dagger}_j  \left(\hat{H} - E_0 \right) \hat{c}_k \hat{c}_l |\Phi_0 \rangle \nonumber \\
           &= -\delta_{ik}\delta_{jl} (\epsilon_{i}+\epsilon_{j}-2\mu)+\langle ij||kl \rangle\, .
	   \label{eq:f-pprpa_CD}
\end{align}
In Eq.~(\ref{eq:f-pprpa_CD}), $\hat{c}^\dagger_p (\hat{c}_p)$ is the creation (annihilation) operator for a single-particle state $p$.
Due to the symmetry properties of the above integrals,
within losing generality, the orbital indices can be restricted to $i<j, k<l$, and  $a<b, c<d$. The indices of matrices $C$ and $D$ correspond to
particle pairs and hole pairs respectively. 
The numbers of particle and hole pairs are
\begin{align}
	N_{pp}& =N_{p}(N_{p}-1)/2\, ,  \nonumber \\
	N_{hh}& =N_{h}(N_{h}-1)/2\, . \nonumber
\end{align}
Consequently, $C$, $D$ are square matrices of dimensionality $N_{pp} \times N_{pp}$ and $N_{hh} \times N_{hh}$ respectively. On the other hand, in contrast with
the phRPA case, $B$ in Eq.~(\ref{Eq:ppRPA_matrix}) becomes a rectangular matrix of $N_{pp} \times N_{hh}$,
\begin{equation}
	\label{eq:pprpa_B}
  B_{ab,ij} = \langle ab|| ij \rangle\, . 
\end{equation}
The eigenvalues $\omega_{n}$ obtained from Eq.~(\ref{Eq:ppRPA_matrix}) are split into two groups depending on their sign: the positive eigenvalues are the
excitation energies of the $(N+2)$-particle system and the negative eigenvalues correspond to (the negative of) excitation energies of the $(N-2)$-particle system. 
The corresponding 
\textit{f}-ppRPA correlation energy can be obtained as 
\begin{equation}
E^{\textit{f}\textnormal{-ppRPA}}_c =\sum_{n}^{N_{pp}}\omega_{n}^{N+2} - Tr\{C\}  \,,
\label{eq:Ec_fppRPA}
\end{equation}
or equivalently 
\begin{equation}
E^{\textit{f}\textnormal{-ppRPA}}_c = -\sum_{n}^{N_{hh}}\omega_{n}^{N-2} - Tr\{D\} \,.
\label{eq:Ec_fppRPA_1}
\end{equation}
It is worthwhile to point out that in ppRPA the inclusion of chemical potential $\mu$ in the definition of matrices $C$ and $D$ is 
not strictly necessary, and the eigenvectors and the final ppRPA correlation energy are not affected by the $\mu$ value. However, it is convenient to do so 
since then the obtained eigenvalues $\omega_n$ can be naturally grouped into positive modes and negative modes, with clear physical meanings as stated above.

\begin{figure}[ht]
    \centering
        \includegraphics[width=0.45\textwidth]{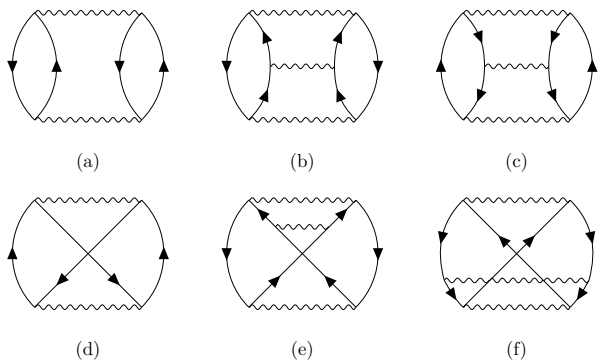} 
	\caption{Goldstone diagrams of second [graphs (a) and (d)] and third orders [graphs (b), (c), (e), and (f)] for ppRPA, among which 
	 the graphs in the upper row [(a), (b), (c)]} correspond to the $d$-ppRPA.
  \label{fig:ppRPA_diagram}
\end{figure}

In analogy to the $d$-phRPA, the \textit{d}-ppRPA is defined by 
solving Eq.~(\ref{Eq:ppRPA_matrix}) with the following definition of $C$,  $D$, $B$ matrices,\\
 \begin{align}
 \label{Eq:dpp_submatrices}
 C_{ab,cd}  & =\delta_{ac}\delta_{bd} (\epsilon_{a}+\epsilon_{b}-2\mu)+\langle ab|cd \rangle  \nonumber \\ 
 D_{ij, kl} &=-\delta_{ik}\delta_{jl} (\epsilon_{i}+\epsilon_{j}-2\mu)+\langle ij|kl \rangle \\ \,
 B_{ab, ij} &=\langle ab| ij \rangle \nonumber 
 \end{align}
 without antisymmetrizing the two-electron Coulomb integrals. Here, it is important to note that, in contrast with \textit{f}-ppRPA, 
 for \textit{d}-ppRPA the orbital index restrictions ($i<j$, $k<l$ and $a<b$, $c<d$) should not be imposed any longer, due to the loss of antisymmetry. 
 The \textit{d}-ppRPA correlation energy expression is the same as
 the \textit{f}-ppRPA case [Eq.~(\ref{eq:Ec_fppRPA}) or (\ref{eq:Ec_fppRPA_1})], but $\omega_n$ should be the \textit{d}-ppRPA eigenvalues 
 and $C$/$D$ matrices should be the non-antisymmetrized integrals as defined in Eq.~(\ref{Eq:dpp_submatrices}). In Fig.~\ref{fig:ppRPA_diagram} we present
 the Goldstone diagrams \cite{Szabo/Ostlund:1989} of second and third orders for $ppRPA$. These diagrams in general have a ladder structure, but those in
 the upper row with the two legs (particle or hole lines) closed into themselves are \textit{direct} diagrams and graphically represent 
 the $d$-ppRPA introduced in this work.


The leading contribution in both ppRPA and phRPA is their corresponding second-order terms, which are the same for the two RPA channels, as
can be easily seen from the corresponding Goldstone diagrams.  The second-order correlation energy in \textit{full} phRPA or \textit{full} ppRPA 
is equivalent to the M{\o}ller-Plesset second order perturbation theory (MP2) \cite{Moller/Plesset:1934}, if the Hartree-Fock reference is used. 
In this work, we term the second-order correlation energy as MP2 for convenience, even if the reference state is not obtained from the Hartree-Fock theory.
In the same language of phRPA and ppRPA discussed above, we shall term the original MP2 as \textit{full} MP2 ($f$-MP2), whose correlation energy is given by
  \begin{equation} \label{eq:Ec_fMP2}
  \begin{split}
   E^{\textit{f}\text{-MP2}}_{c}= -\dfrac{1}{2}\sum_{ijab}\dfrac{\langle ij|ab \rangle \langle ab||ij \rangle}{\epsilon_{a}+\epsilon_{b}-\epsilon_{i}-\epsilon_{j}}
	  \, .
  \end{split}
  \end{equation}
  By contrast, the \textit{direct} MP2 (\textit{d}-MP2) correlation energy is obtained as\,  \\
   \begin{equation} \label{eq:Ec_dMP2}
  \begin{split}
   E^{\textit{d}\text{-MP2}}_{c}= -\dfrac{1}{2}\sum_{ijab}\dfrac{\langle ij|ab \rangle \langle ab|ij \rangle}{\epsilon_{a}+\epsilon_{b}-\epsilon_{i}-\epsilon_{j}}\, . 
  \end{split}
  \end{equation}

  Both phRPA and ppRPA contain a subset of diagrams of the coupled cluster double (CCD) theory 
  \cite{Cizek:1966,Bishop/Luehrmann:1978,Scuseria/Henderson/Sorensen:2008,Scuseria/Henderson/Bulik:2013,Peng/Steinmann/etal:2013}. 
  While both channels of RPA show some promising performance, they also have some known drawbacks. As discussed above,
  one of the {motivations} of the present work is to check if it is possible to combine them to arrive at a better theory, while not going to the full CCD method.
  A straightforward way to combine phRPA and ppRPA is to add them up, while subtracting the double-counted MP2 term,
 \begin{equation} \label{Eq:combined_rpa}
{ E}^\text{comb-RPA}_{c}={E}^\text{ppRPA}_{c}+{E}^\text{phRPA}_{c}-E^\text{MP2}_{c}\, .
\end{equation}  \\
 In this combined RPA (denoted as ``comb-RPA" in the following) scheme, all three correlation energies appearing on the right-hand side are 
 obtained either in their \textit{direct} or \textit{full} flavor, but not mixing up the two flavors. The combination of the $f$-phRPA and $f$-ppRPA, with
 $f$-MP2 subtracted, is termed as $f$-comb-RPA; analogously, the combination of the $d$-phRPA and $d$-ppRPA, with  
  $d$-MP2 subtracted, is termed as $d$-comb-RPA.  We emphasize that both \textit{f}-comb-RPA and \textit{d}-comb-RPA
 are free of double-counting effects at all orders.

 We would like to point out that the combination scheme defined in Eq.~(\ref{Eq:combined_rpa}), despite being free of double counting, is not derived 
  rigorously from more fundamental theories. It should rather be viewed as an empirical ansatz whose performance needs to check \textit{a posteriori}. One may
  also design alternative double-counting-free schemes, .e.g., by simply averaging phRPA and ppRPA, or by combining the two RPA flavors in a range-separation 
  framework.  In a pioneering work by Shepherd, Henderson, and Scuseria \cite{Shepherd/Henderson/Scuseria/:2014} short-range ppRPA and long-range phRPA are combined. 
  Initial tests of their scheme for homogeneous electron gas show promising performance.
 In this connection, it is also interesting to compare Eq. (\ref{Eq:combined_rpa}) to the quasiparticle RPA (qp-RPA) scheme of Scuseria, Henderson, 
  and Bulik \cite{Scuseria/Henderson/Bulik:2013} which consists in a simple summation of ppRPA and phRPA, without eliminating the double-counted MP2 term.
  It should also be noted that, in qp-RPA a prefactor of $1/2$ instead of $1/4$ is used for \textit{f}-phRPA. 

 \section{Computational details}
 \label{sec:comput}
 Both the \textit{direct} and \textit{full} ppRPA equations are implemented within the all-electron, numerical atomic orbital (NAO) based computer code package 
 FHI-aims \cite{Blum/etal:2009,Havu/etal:2009,Ren/etal:2012}. FHI-aims primarily employs NAOs as basis functions to expand molecular orbitals,
 but, if needed, {Gaussian}-type orbitals (GTOs) can also be used for comparison purpose. The \textit{direct} phRPA has already been
 implemented in FHI-aims \cite{Ren/etal:2012}, based on the resolution of identity (RI) technique, together with an integration over the
 imaginary frequency axis. This allows for a relatively efficient $O(N^4)$ scaling of \textit{direct} phRPA calculations. However, such 
 a low-scaling algorithm cannot be applied to the \textit{full} phRPA.  In this work, we implemented Eq.~(\ref{eq:Ec_fphRPA}) and Eq.~(\ref{eq:Ec_dphRPA}) 
 straightforwardly, which scales as $O(N^6)$, to obtain both the \textit{full} and \textit{direct} phRPA correlation energies. In the $d$-phRPA case, 
 the obtained results agree with the previous 
 RI-based implementation to a high precision. 
 The \textit{full} ppRPA was implemented in FHI-aims following the work of Yang \textit{et al.}  \cite{Yang/Aggelen/Steinmann/etal:2013}. The implementation of 
 \textit{direct} ppRPA is straightforward by replacing the $C$, $D$, $B$ matrices defined in Eqs.~(\ref{Eq:ppRPA_matrix}) and (\ref{eq:pprpa_B}) by those defined 
 in Eq.~(\ref{Eq:dpp_submatrices}). However, due to the loss of symmetry properties, for $d$-ppRPA the dimension of $C$,$D$, $B$ matrices 
 become $N_{p}^{2}\times N_{p}^{2}$, $N_{h}^{2}\times N_{h}^{2}$, and $N_{p}^{2}\times N_{h}^{2}$ respectively  as in \textit{f}-ppRPA for mixed-spin pair channels \cite{Yang/Aggelen/Steinmann/etal:2013}.
 The restricted/unrestricted Hartree-Fock (RHF/UHF), MP2 \cite{Moller/Plesset:1934,Szabo/Ostlund:1989}, and 
 rPT2 \cite{Ren/etal:2013} have already been implemented in FHI-aims.
  In this work, we also implemented the \text{direct} MP2, as defined in Eq.~(\ref{eq:Ec_dMP2}), in FHI-aims. Our $f$-MP2 and $d$-MP2 calculations can not only 
  be done on top of the HF reference, but also on top of density functional approximations (DFAs).

  In addition to the phRPA and ppRPA results, in this work we will also present results of Hartree-Fock, MP2, rPT2, the coupled cluster theory with 
  singles and doubles substitutions (CCSD) \cite{Cizek:1966,Bartlett/Musial:2007}, 
  and Multi-Reference Configuration Interaction Single Double (MRCISD) \cite{Werner/Knowles:1988} for comparison.
   The Hartree-Fock, MP2, rPT2, various flavors of RPA calculations are
   done with FHI-aims.  The CCSD calculations are done with the Gaussian 09 W package \cite{g09} with the exception of H2 molecule based on UHF 
   [Fig.~\ref{fig:H2_HF_spin_unrestricted}], for which we used the recent implementation of Shen \textit{et al.}\cite{2018arXiv181008142S} in FHI-aims. 
   Finally, MRCISD calculations are done with  the COLUMBUS package  \cite{COLUMBUS}. 

   We examined the binding energy curves of four homo-nuclear dimers, including the covalently bonded hydrogen dimer (H$_2$) 
   and nitrogen dimer (N$_2$), the ionically bonded hydrogen fluoride (HF) dimer, and van der Waals (vdW) bound Argon dimer (Ar$_2$). For the former three dimers,
    the {Gaussian} cc-pVTZ \cite{Dunning:1989} basis sets are used, whereas for Ar$_2$, the aug-cc-pVTZ basis set was used instead. 
    These basis sets admittedly cannot yield converged binding energy curves, but are sufficient for the present purpose, i.e., 
    comparing the qualitative dissociation behavior of molecular dimers obtained by different methods. In Appendix~\ref{app:accuracy}, the numerical settings employed
    in our FHI-aims calculations are presented, together with benchmark results of the $f$-ppRPA total energies for a sequence of atoms, in comparison
    with published results in Ref.~[\onlinecite{Peng/Steinmann/etal:2013}].

 
 \section{Results and discussion}
 \label{sec:results}
We present in this section the calculated results for four closed-shell dimers (H$_2$, N$_2$, HF, and Ar$_2$) and one open-shell dimer (H$_2^{+}$).
The obtained results of all correlated methods depend on the reference state on which they are based. For closed-shell molecules spin-restricted references 
are often used since they provide a stringent test for static correlation errors. However, for stretched bonds spin-unrestricted references can yield lower
energies and better behaved dissociation curves, at the price of broken spatial and spin symmetries. In this section, we mainly use RHF as
the reference for closed-shell dimers. For H$_2$, results of correlated methods based on UHF will also be shown for comparison. For the open-shell molecule
H$_2^{+}$, as a natural choice the UHF reference will be used. For RPA methods, instead of Hartree-Fock, the KS reference is often used in practical
calculations. The influence of preceding reference states obtained with different functionals will
be illustrated in the Appendix~\ref{app:pbe_ref} by presenting binding energy curves for H$_2$ and Ar$_2$ based on 
the generalized gradient approximation of Perdew, Burke, and Ernzerhof (PBE) \cite{Perdew/Burke/Ernzerhof:1996}. Basis set superposition errors are
not corrected in the presented results, since in this work we are focused on the relative trends of different methods.

\subsection{H$_2$ and H$_2^{+}$}
 The dissociation curves of the H$_2$ and H$_2^{+}$ are of particular interest for testing DFAs and/or quantum chemistry methods.
 It appears that none of the current DFAs are able to satisfactorily describe the dissociation behavior of both dimers \cite{Cohen/etal:2012}.  
 Therefore{,} H$_2$ and H$_2^{+}$ dimers are the most important target systems for the recent efforts on designing novel electronic-structure 
 methods \cite{IgorZhang/etal:2016a,IgorZhang/etal:2016b}. 
 Our purpose here is to examine the behavior of both the \textit{direct} and \textit{full}
 RPA methods, and especially the combination of ppRPA and phRPA (comb-RPA) as defined in Eq.~(\ref{Eq:combined_rpa}), for these two dimers.


\begin{figure}[ht]
   \begin{minipage}{0.46\textwidth}
        \includegraphics[width=\textwidth]{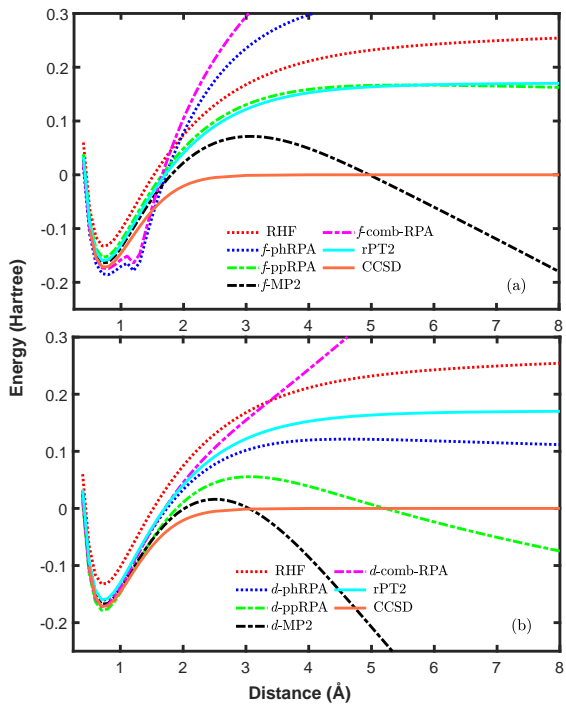} 
   \end{minipage}
  \caption{Binding energy curves for $\text{H}_{2}$ with the RHF reference and cc-pVTZ basis set. Panel (a): RPA and MP2 calculations
           are done with exchange contributions included (\textit{full} RPA and MP2); 
           panel (b): RPA and MP2 calculations are done without including
           exchange contributions (\textit{direct} RPA and MP2). RHF, rPT2, and CCSD results are included in both panels for comparison. }
  \label{fig:H2}
\end{figure}


The binding energy curves of H$_2$ obtained from various methods, based on the RHF reference, are presented in Fig.~\ref{fig:H2}. In panel (a),
 the results of the \textit{full} 
RPA methods, including $f$-ppRPA, $f$-phRPA, and $f$-comb-RPA, are presented, whereas in panel (b), the results from the corresponding 
\textit{direct} RPA methods are presented. The $f$-MP2 and $d$-MP2 results are also included 
in panel (a) and (b) respectively. In addition, the rPT2 and CCSD results are presented in both panels for comparison. Note that the latter two methods are 
treated here as they are originally, without being further separated into \textit{direct} and \textit{full} flavors. Especially the CCSD results are exact for 
one- and two-electron systems, and hence should be regarded as the reference here. In this case, since
the Hartree-Fock reference is used, the rPT2 method reduces to RPA+SOSEX\cite{Grueneis/etal:2009,Paier/etal:2010}. To highlight the influence of the preceding
functionals on the obtained results, in the Appendix \ref{app:pbe_ref}, the same plots, albeit with the PBE reference, are presented (cf. Fig.~\ref{fig:H2_pbe}).

All RPA methods describe the H$_2$ molecule satisfactorily around the equilibrium bond length. For stretched H$_2$, the \textit{f}-phRPA becomes unstable
around the Coulson-Fisher point \cite{Coulson/Fischer:1949} and the binding energy curve has a cusp for the bonding distance around 1.3 \AA{}. Compared to 
the \textit{direct} RPA methods, the corresponding \textit{full} RPA methods are more repulsive, indicating that the exchange contributions give rise to 
a positive contribution to the binding energies of the H$_2$ dimer. 
All RPA methods, except for $d$-ppRPA, are too repulsive for the stretched H$_2$. Namely, they yield an energy that is too high for the stretched H$_2$ dimer, similar to the behavior of RHF.  Only at the dissociation limit (not shown here), $d$-phRPA and $f$-ppRPA have been shown to yield the correct total energy (i.e., twice of the energy of an isolated H atom) \cite{Fuchs/Gonze:2002,Hesselmann/Goerling:2011b,Aggelen/etal:2013}, although the asymptotic behavior of these methods is still
incorrect. 
In contrast, MP2 yields an energy that is too low for the stretched dimer,  and the MP2 energy becomes diverging in the dissociation limit. At the
intermediate bonding distance, the $d$-MP2 is more negative than the $f$-MP2, indicating again that the exchange contribution is positive.
The behavior of $d$-ppRPA is somewhat similar to MP2, giving a binding energy that is too low at large bonding distances.
Putting phRPA and ppRPA together, the comb-RPA defined in Eq.~(\ref{Eq:combined_rpa}) leads to even more repulsive binding energies for stretched H$_2$, as can be
seen in Fig.~\ref{fig:H2}(a,b) for both the \textit{full} and \textit{direct} RPA flavors. 
Mathematically, by subtracting the very negative MP2 total energy, it comes out naturally that the resultant comb-RPA total energies are much too high. 
Also, the instability problem of $f$-phRPA around the Coulson-Fisher point is inherited by the $f$-comb-RPA result. In summary, although the comb-RPA scheme as 
defined in Eq.~(\ref{Eq:combined_rpa}) is free of double-counting,  it unfortunately does not work 
in practice for molecular dissociations, in both \textit{full} and \textit{direct} flavors of RPA schemes, if one insists using spin-restricted references. 
Furthermore, from this comparative study, it is also clear that the phRPA behaves better in its \textit{direct} flavor, whereas the ppRPA behaves better 
in its \textit{full} flavor. The $d$-phRPA and $f$-ppRPA are indeed the usual choice in the literature.

The situation is however quite different if the UHF reference is used. The UHF binding energy curve coincides with the RHF one around the equilibrium distance;
beyond the Coulson-Fischer point, the UHF energy is consistently lower, and follows correct asymptotic behavior towards the dissociation limit. On top of UHF,
all correlated methods can produce the correct dissociation limit of H$_2$.  However, the $f$-phRPA binding energy curve displays a well-known cusp around
the Coulson-Fischer point and this pathological behavior carries over to the corresponding $f$-comb-RPA. Other RPA schemes, as well as MP2 and rPT2, don't
suffer from this problem. However, the binding energy curves from these methods appear to decay faster to zero compared to the reference CCSD@UHF reference.
Note that for H$_2$, the results from CCSD@RHF and CCSD@UHF are identical, and both are exact. 
\begin{figure}[H]
    \centering
    \begin{minipage}{0.46\textwidth}
        \includegraphics[width=1.\textwidth]{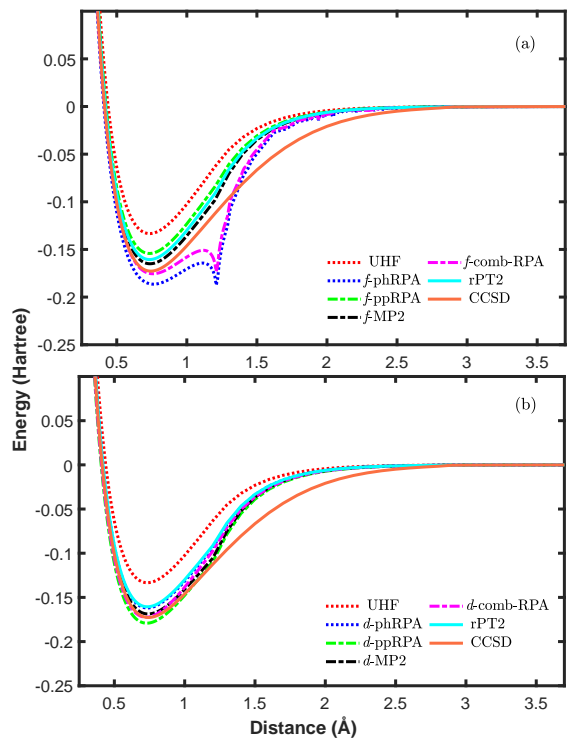} 
    \end{minipage}
        \caption{Binding energy curves for H$_2$ with the UHF reference and cc-pVTZ basis set. Panel (a): \textit{full} RPA/MP2 results; panel
          (b): \textit{direct} RPA/MP2 results. The plots here differ from Fig.~\ref{fig:H2} only that the UHF reference instead of RHF is used.} 
  \label{fig:H2_HF_spin_unrestricted}
\end{figure}

  \begin{figure}[h]
    \centering
    \begin{minipage}{0.46\textwidth}
        \includegraphics[width=\textwidth]{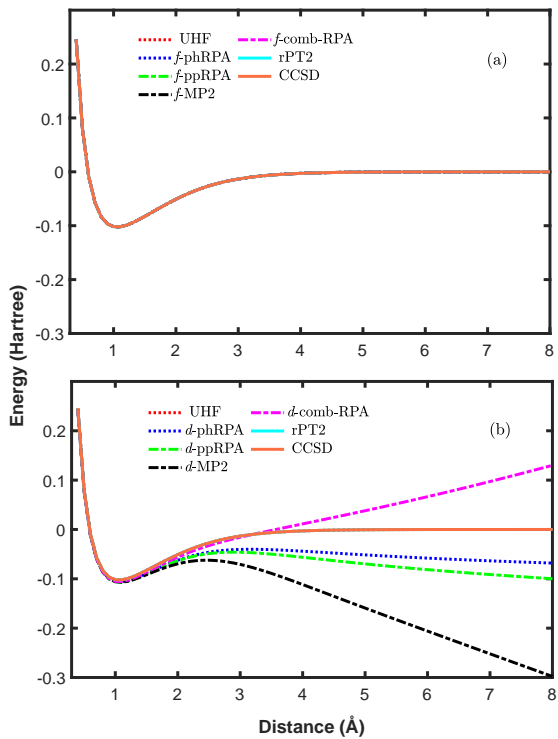} 
    \end{minipage}
        \caption{Binding energy curves for $\text{H}_{2}^{+}$ with the UHF reference and cc-pVTZ basis set. 
             Panel (a): RPA and MP2 calculations
             are done with exchange contributions included (\textit{full} RPA and MP2); panel (b): RPA and MP2 calculations are done without including
             exchange contributions (\textit{direct} RPA and MP2). UHF, rPT2, and CCSD results are included in both panels for comparison. }
  \label{fig:H2+}
\end{figure}
Similar comparative studies for the dissociation of the H$_2^{+}$ molecule are presented in Fig.~\ref{fig:H2+}. Now the system contains only one electron, and the 
UHF method is exact and is taken as the reference here. All other methods are based on the UHF reference state.
Besides the UHF,  CCSD and rPT2 are also exact for H$_2^{+}$ in theory. Moreover, all the \textit{full} RPA schemes, as well as MP2, are able to produce the correct 
dissociation curves for H$_2^{+}$, as can be seen in Fig.~\ref{fig:H2+}(a). The incorporation of the exchange contributions cancels the one-electron self-correlation 
energy, and consequently the \textit{full} RPA schemes correctly yield zero correlation energy for H$_2^{+}$. In contrast, both the $d$-phRPA and 
$d$-ppRPA yield too low energy for stretched H$_2^+$,  indicating the presence of strong charge delocalization errors in these two approaches. In fact,
the problem is even more severe for $d$-MP2, which yields
even more negative (and eventually diverging) energy than \text{direct} RPA's. As a consequence, the $d$-comb-RPA binding energy curve 
becomes much too repulsive for heavily stretched H$_2^{+}$. 

In this context, we would like to briefly comment on the delocalization error of the RPA methods. As first pointed out
by Mori-S\'anchez, Cohen, and Yang \cite{Mori-Sanchez/etal:2012}, the $d$-phRPA suffers from severe delocalization errors, as 
manifested in the dissociation of H$_2^+$. Here we can see this problem arises from the presence of the artificial one-electron self-correlation energy in $d$-phRPA. 
In fact, for ppRPA, when the exchange contributions are taken out, the $d$-ppRPA also suffers from this error, though to a less extent. We note that, for H$_2$, 
such self-correlation errors are still present in \textit{direct} RPA. However, the negative self-correlation energy becomes advantageous for H$_2$ since it cancels the positive RHF energy at the dissociation
limit. This is in line with the analysis of Henderson and Scuseria \cite{Henderson/Scuseria:2010} that the self-interaction errors of \textit{direct} phRPA mimics the
static correlation effect. The high RHF energy for stretched H$_2$ stems from the on-site Coulomb interaction of two-electrons occupying the same atom --
a ``double-occupation" (or ionic) configuration 
that is unavoidable when a single-determinant description of stretched H$_2$ is employed. Such a high RHF energy arising from the unphysical ionic configurations
was compensated by the negative RPA correlation energy of the magnitude at the dissociation limit. However, since this compensation is not complete except at the
infinite separation, and the asymptotic behavior of $d$-phRPA@RHF is still incorrect for dissociating H$_2$. In summary, $d$-phRPA correlation part itself 
does not really behave differently for
the dissociation of H$_2$ or H$_2^+$. It is the different behavior of the preceding Hartree-Fock calculation in these two systems that leads to an overall
drastically different performance of the $d$-phRPA scheme for these two systems.

\subsection{N$_2$}
 \begin{figure}[h]
    \centering
    \begin{minipage}{0.46\textwidth}
        \includegraphics[width=1.\textwidth]{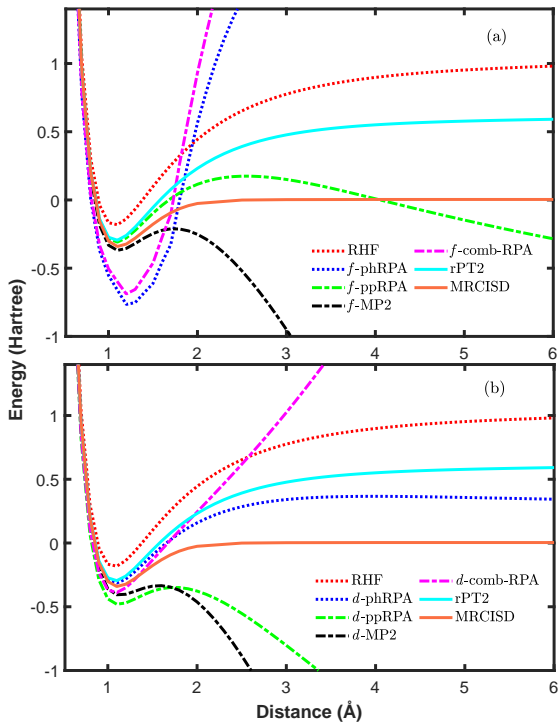} 
    \end{minipage}
	 \caption{Binding energy curves for $\text{N}_{2}$ with the RHF reference and cc-pVTZ basis set. Panel (a): \textit{full} RPA/MP2 results; panel
	  (b): \textit{direct} RPA/MP2 results. The curves are presented in the same way as Fig.~\ref{fig:H2}, except that the reference curve is now given by MRCISD.}
 \label{fig:N2}
\end{figure} 
Now we look at the triply-bonded nitrogen dimer (N$_2$).  Correctly describing the dissociation behavior of N$_2$ is a long-standing challenge for
any single-reference electronic-structure method. Indeed, for N$_2$ even the CCSD and CCSD(T) methods do not work, and here we use the results obtained by 
the multi-reference configuration interaction with single and double excitations (MRCISD) as the reference here.  In Fig.~\ref{fig:N2} we plotted the 
binding energy curves obtained by \textit{full} and \textit{direct} RPA methods (based on the RHF reference) respectively in panel (a) and (b).
The \textit{full} and \textit{direct} MP2, as well as the rPT2 results are also plotted for comparison. For N$_2$, the different
methods already yield quite different binding energies around the equilibrium distance. The $d$-phRPA was known to be able to produce the correct
dissociation limit for N$_2$ \cite{Furche:2001}, although a positive bump was formed at intermediate bond lengths.
Comparing the results of $f$-phRPA to $d$-phRPA, we see that 
the exchange terms increase the bond strength (more attractive) around the equilibrium distance while weakening it (more repulsive)
at the intermediate and large bonding distances. This results in a very steep binding energy curve with large curvature. Such an (unphysical) feature
also carries over to  {the} $f$-comb-RPA scheme, as can be seen from Fig.~\ref{fig:N2}.  The $f$-ppRPA performs well around the equilibrium region, 
but forms a positive bump at the intermediate bonding distances, and eventually goes below the energy zero at large bond lengths. 

Comparing panel (a) to panel (b) reveals that the exchange terms{,} in general{,} makes the binding energy curve
more repulsive at intermediate and large bonding distances, not only for phRPA, but also for ppRPA and MP2. Without including exchange contributions, 
the binding energies of $d$-ppRPA behaves similar to MP2, and falls below the energy zero already at intermediate bonding distances. 
The rPT2 (i.e., $d$-phRPA+SOSEX here) result for N$_2$ behaves similarly to $d$-phRPA and $f$-ppRPA
at the equilibrium distance, but saturates at too high energies in the dissociation limit. Similar to the H$_2$ case, the comb-RPA schemes again performs
very badly for N$_2$. The rising of the $f$-comb-RPA binding energy curve is even steeper than $d$-comb-RPA for increasing bond lengths, arising from 
a similar behavior of $f$-phRPA.

\subsection{HF}
\begin{figure}[h]
    \centering
    \begin{minipage}{0.46\textwidth}
        \includegraphics[width=1.\textwidth]{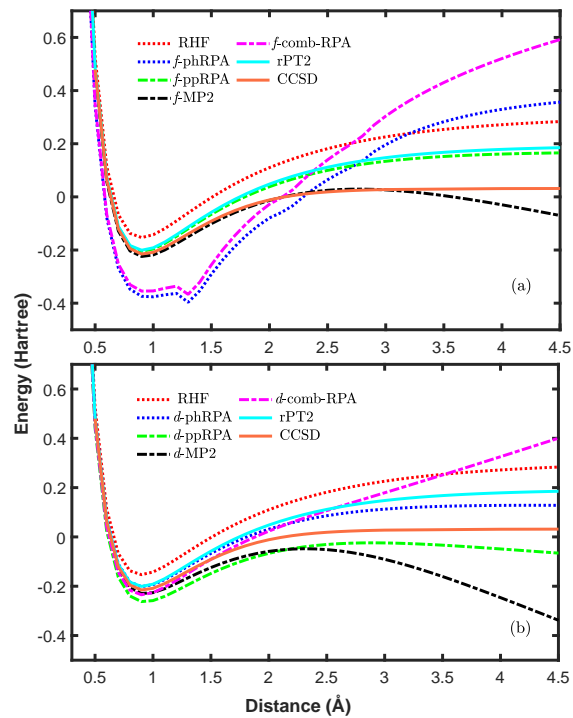} 
    \end{minipage}
	\caption{Binding energy curves for hydrogen fluoride ($\text{HF}$) with the RHF reference and cc-pVTZ basis set. Panel (a): \textit{full} RPA/MP2 results; 
	 panel (b): \textit{direct} RPA/MP2 results. The curves are presented in the same way as Fig.~\ref{fig:H2}.} 
\label{fig:HF}
\end{figure} 
Next, we examine a dimer of ionic character -- the hydrogen fluoride (HF). In Fig.~\ref{fig:HF}, the RHF, RPA, MP2, and rPT2 results for the HF dimer are presented.
Now the CCSD result is not exact for the HF dimer, but still provides a high-quality reference. Similar to the N$_2$ case, the $f$-phRPA already overbinds 
the HF dimer around the equilibrium distance, suffers from instabilities at bonding distances around 1.4 \AA, and gets too repulsive for large bond lengths. 
Such these behaviors carry over to $f$-comb-RPA. Surprisingly, MP2 performs pretty well at the equilibrium and intermediate bonding distances, but then drops down
(and eventually diverging) at large distances. The $f$-ppRPA also performs well around the equilibrium distance, and follows closely the (too repulsive) rPT2 curve 
for large bond lengths. Removing the exchange contributions, the $d$-phRPA behaves much more reasonably over a wide range of bonding distances. 
On the other hand, the $d$-ppRPA curve becomes attractive at large bonding distances , and falls below the CCSD curve. 
The $d$-comb-RPA curve behaves similarly to the $d$-phRPA around the equilibrium distance, but gets too repulsive for large distances, 
arising from the opposite behavior of $d$-MP2.
\subsection{Ar$_2$}
 \begin{figure}[h]
    \centering
    \begin{minipage}{0.46\textwidth}
        \includegraphics[width=1.\textwidth]{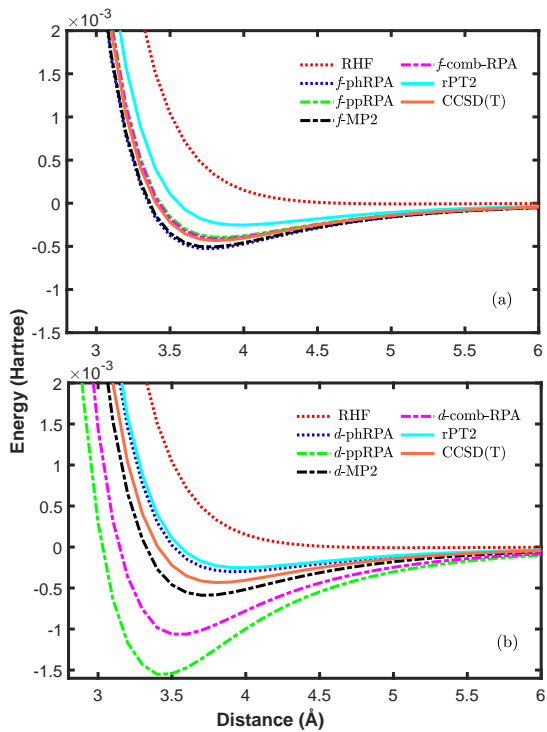} 
    \end{minipage}
	 \caption{Binding energy curves for  Ar$_2$ with the RHF reference and gaussian aug-cc-pVTZ basis set.  Panel (a): \textit{full} RPA/MP2 results;
		  panel (b): \textit{direct} RPA/MP2 results. The curves are labeled in the same way as Fig.~\ref{fig:H2}, except that the reference curve
		  is now given by CCSD(T). The frozen-core approximation is used in all correlated calculations. }
  \label{fig:Ar2}
\end{figure} 
Finally, we look at a prototypical dimer bound purely by dispersion interactions -- Ar$_2$. The results from various methods are presented in 
Fig.~\ref{fig:Ar2} (a) and (b). One can see that the $f$-ppRPA performs rather well for Ar$_2$, yielding a binding energy curve that follows closely 
the CCSD(T) reference curve. In contrast with the covalently and ionically bonded dimers, the $f$-phRPA does not exhibit any pathological behavior for Ar$_2$.
This is because the Ar atom itself has a closed-shell electronic structure, and the RHF solution is stable for Ar$_2$ for all inter-atomic distances. 
As a consequence, MP2 is well-behaved for Ar$_2$, although a well-known overbinding behavior can be noticed. In this case,
the $f$-phRPA result closely resembles that of (\textit{full}) MP2, and putting all three ingredients together, the $f$-comb-RPA performs
remarkably well, producing a binding energy curve that is highly accurate. 

Remarkably, in the case of Ar$_2$, the exchange contributions seem to have an opposite effect on phRPA and ppRPA. Without including the exchange contributions,
the $d$-phRPA curve become more repulsive, showing a well-known underbinding behavior for Ar$_2$. On the contrary, the $d$-ppRPA curve becomes much more attractive,
vastly overbinding Ar$_2$. As a consequence, in contrast to $f$-comb-RPA, the $d$-comb-RPA overbinds the Ar$_2$ dimer significantly. We note that, 
the rPT2 underbinds Ar$_2$ in Fig.~\ref{fig:Ar2}; this is because rPT2 here is based on the RHF reference, and the renormalized singles contribution is not included. 
The rPT2@PBE scheme instead yields an accurate binding energy curve for Ar$_2$, as can be seen from Fig~\ref{fig:Ar2_pbe} in the Appendix~\ref{app:pbe_ref}.


\section{Conclusion}
\label{sec:conclusion}
In this work, we implemented the ppRPA scheme within the all-electron, NAO-based code package -- FHI-aims. Benchmark calculations show that our implementation, 
based on
the resolution of identity approximation to the two-electron Coulomb integrals, agrees remarkably well with the previous implementation of Peng and coauthors
\cite{Peng/Steinmann/etal:2013}. We performed a systematic comparative study of the 
behavior of the ppRPA and phRPA for describing the dissociation of diatomic molecules of different bonding characters. The novel aspect in our research is
that we introduced a \textit{direct} ppRPA, whereby the exchange contributions are excluded from the formalism, in a similar fashion as done in the
\textit{direct} phRPA. This allows us to compare phRPA and ppRPA on an equal footing, separately for \textit{direct} and \textit{full} RPA flavors. 
While benchmark calculations show that both phRPA and ppRPA are not able to dissociate correctly all types of dimers, but generally speaking, the
phRPA is better employed in its \textit{direct} flavor, without including the exchange terms (i.e., $d$-phRPA), while the \textit{full} RPA is better employed in 
its \textit{full} flavor, with exchange terms included (i.e., $f$-ppRPA). In this work, we also pointed out the seemingly different performance of 
$d$-phRPA for H$_2$ and H$_2^{+}$ mainly arises from the preceding RHF/UHF for H$_2$ and H$_2^{+}$, and not from the $d$-phRPA correlation energy itself. 

In an attempt to combine both phRPA and ppRPA, we examined a simple procedure (Eq.~\ref{Eq:combined_rpa}), whereby the phRPA and ppRPA correlation energies 
are added together, with the double-counted second-order (MP2) correlation energy removed. This scheme, although containing no double-counting terms, 
yields worse and often unphysical results for the dissociation of covalent and ionic diatomic molecules. The behavior stems from the bad performance of MP2 
for describing
stretched molecules, but is manifested in an opposite way, resulting in too repulsive binding energy curves for large bonding distances. In the quasi-particle RPA 
scheme examined by Scuseria \textit{et al.} \cite{Scuseria/Henderson/Bulik:2013}, the phRPA and ppRPA correlation energies are summed up, but the doubly-counted MP2 
terms are not excluded. This quasiparticle RPA scheme does not suffer from some of the drastic failures of the comb-RPA scheme examined in this work, but
vastly overestimates the total and binding energies of molecular systems. Scuseria \textit{et al.} \cite{Scuseria/Henderson/Bulik:2013} attributed this failure
to the neglecting of the inter-channel coupling terms between phRPA and ppRPA. 

The pertinent question is if it is possible to develop a theory between RPA and CCSD, which is close to RPA in the computational cost, but close to CCSD in
the accuracy? The present work shows that it is highly nontrivial to achieve this goal. A straightforward combination of the ph and pp channels
of RPA does not work. More investigations along these lines are required to answer this question.

\section*{Acknowledgments}
XR acknowledges the support from Chinese National Science Foundation (No. 11574283, 11874335), the Max Planck Partner Group project, and the Fundamental 
Research Funds for the Central Universities. We thank Yang Yang and Meiyue Shao for helpful discussions regarding the ppRPA implementation, Igor Ying Zhang for 
the discussion of the Ar$_2$ results, as well as Hans Lischka and Thomas M{\"u}ller for providing the COLUMBUS code.  The numerical calculations have been partly 
done on the USTC HPC facilities.

\begin{appendix}
\section{Numerical accuracy of our ppRPA implementation in FHI-aims}
\label{app:accuracy}

%
As mentioned above, in this work we are concentrating on the qualitative features of different computational schemes, rather than presenting highly converged 
	results with respect to the basis set size.  To facilitate a direct comparison with literature results,  we employed Gaussian basis sets, 
	which are however treated numerically in FHI-aims \cite{Blum/etal:2009}. 
	The real-space integration are done by a summation over atom-centered overlapping numerical grid.
	The accuracy of the numerical integration is controlled by two parameters of the spherical grids positioned around each atom: 
	the number of radial integration shells $N_r$ and the angular grid points in the outermost radial shell (denoted below as \textit{outer\_grid}). 
	Furthermore, the two-electron Coulomb repulsion integrals (ERI) are evaluated using the resolution-of-identity (RI) approximation. 
	In FHI-aims, the auxiliary basis functions (ABFs) used in the RI expansion are generated from ``on-site" products of atomic orbitals, and the redundancy of
	such products are further eliminated through the Gram-Schmidt orthonormalization procedure \cite{Ren/etal:2012,Ihrig/etal:2015}. The accuracy of RI 
	is affected by the number of ABFs, which is in turn controlled by the threshold $\eta$ (keyword \textit{prodbas\_acc} in FHI-aims) in the Gram-Schmidt 
	procedure, and by the threshold $\theta$ (keyword \textit{prodbas\_threshold}) set for singular value decomposition (SVD) of the Coulomb matrix (
	in order to invert it) within 
	the ABFs.  Thus the numerical accuracy of RPA total energies in FHI-aims, for a given set of single-particle atomic orbitals, are affected both by the 
	numerical integration grid and by the RI accuracy.

	In Table~\ref{tab:grid_performance} we present $f$-ppRPA total energies (more precisely the deviation from the reference value) for the N$_2$ molecule
	for a set of successively denser integration grid. From Table~\ref{tab:grid_performance}, one can see that, for $rm>=4$, the error incurred by
	numerical integration is below 1 meV for the $f$-ppRPA total energy of N$_2$. Similar accuracy can be achieved for other types of RPA calculations.
	Table~\ref{tab:RI_accuracy} demonstrates the influence of two key parameters involved in the RI approximation on the $f$-ppRPA total energy. One 
	can see that the results here is not much sensitive to the $\eta$ parameter for $\eta >= 10^{-2}$, but an appreciable dependence on the $\theta$ parameter
	is observed. From $\eta<= 10^{-4}$, an accuracy better than 1 meV in the RPA total energy for N$_2$ can be achieved.
\begin{table}  
	\caption{\label{tab:grid_performance} Deviations (in meV) of \textit{f}-ppRPA@RHF total energy for N$_2$ from the reference value due to the real-space 
	 integration grid governed by the number of radial shells $N_r$ and the angular grid points of the outermost shell $outer\_grid$. 
	 In FHI-aims \cite{Blum/etal:2009},
	 $N_r = rm * N_r^0$, where $rm=1, 2, 4, 6, \cdots$ is the so-called \textit{radial multiplier} and $N_r^0=46$ for the N element. 
	The total energy $E=-2972.67642407$ eV, obtained with $rm=6$~$(N_r=276)$ and $outer\_grid=1202$, is taken as the reference value here. 
	The cc-pVTZ basis set is used
	in the calculation. The parameters $\eta= 10^{-2}$ and $\theta=10^{-5}$ are used for used for the RI decomposition of ERIs. }
 \begin{tabular}{|c|ccccc|}
 \hline\hline
	 \diagbox{$rm(N_r)$}{$outer\_grid$}  &  434  & 590 & 770 & 974 &   
	   1202\\   
     \hline 
	 $2 (92)$  &  ~2.559  &  ~2.588 &  ~2.554  &  ~2.559 & ~2.562 \\
	 $4 (184)$ &  ~0.653  &  ~0.669 &  ~0.652  &  ~0.653 & ~0.654 \\
	 $6 (276)$ &  -0.002  &  ~0.014 &  -0.002  &  -0.001 & ~0.000 \\
\hline\hline
\end{tabular}
\end{table}

\begin{table}  
	\caption{\label{tab:RI_accuracy} Deviations (in meV) of the $f$-ppRPA@RHF total energy from the reference value due to the RI approximation. Here
	$\eta$ (\textit{prodbas\_acc}) and $\theta$ (\textit{prodbas\_threshold}) are respectively the thresholding parameter for the Gram-Schmidt 
	orthonormalization and for the SVD decomposition. The total energy $E=-2972.67648033$ eV, obtained with $\eta=10^{-4}$ and $\theta=10^{-6}$ 
	is taken as the reference here. In all calculations $rm=6$ and \textit{outer\_grid}=770 are used for the real-space grid integration.  }
 \begin{tabular}{|c|cccc|}
 \hline\hline
	 \diagbox{$\eta$}{$\theta$}  & $10^{-3}$ &  $10^{-4}$ & $10^{-5}$& $10^{-6}$\\   
     \hline 
	 $10^{-2}$   &  8.821  &   0.416  & 0.058  &  0.052 \\
	 $10^{-3}$   &  7.234  &   0.212  & 0.152  &  0.004 \\
	 $10^{-4}$   &  6.636  &   0.295  & 0.046  &  0.000 \\
\hline\hline
\end{tabular}
\end{table}


 Finally, to validate our ppRPA implementation in FHI-aims, in Table~\ref{tab:pprpa_energy} we present our calculated all-electron $f$-ppRPA 
 total energies, on top of the UHF reference, for a selected set of atoms. Our results obtained using FHI-aims are compared 
 to those of Peng \textit{et al.}\cite{Peng/Steinmann/etal:2013}. The ground-state UHF total energies are also presented for comparison.
 For these calculations, we set $rm=6$ and $outer\_grid = 770$ for the grid integration, and $\eta = 10^{-4}$ and $\theta = 10^{-5}$ for the RI approximation.
 We see that the differences in both UHF and $f$-ppRPA total energies between our implementation and that of Peng \textit{et al.}\cite{Peng/Steinmann/etal:2013}
 are vanishingly small -- only noticeable at the micro-Hartree ($\mu$Ha) level. Such $\mu$Ha-level error in $f$-ppRPA total energies indicates that our RI-based 
 ppRPA implementation is highly accurate.
\begin{table*}  
	\caption{\label{tab:pprpa_energy}UHF and $f$-ppRPA@UHF total energies (in Hartree) for a series of atoms. Results obtained in this work are compared to the literature values of Peng \textit{et al.} \cite{Peng/Steinmann/etal:2013}. The {Cartesian Gaussian} cc-pVTZ basis set is used in both works. The ``Difference" columns present the UHF/$f$-ppRPA total energy differences between this work and Ref.~[\onlinecite{Peng/Steinmann/etal:2013}].}
 \begin{tabular}{crrrcrrr}
 \hline\hline
   \multirow{2}{*}{Atom}  &  \multicolumn{3}{c}{HF} & &  \multicolumn{3}{c}{$f$-ppRPA} \\   
     \cline{2-4} \cline{6-8} 
	 & Ref. \cite{Peng/Steinmann/etal:2013} & This work & \multicolumn{1}{c}{Difference} & & Ref. \cite{Peng/Steinmann/etal:2013} & This work & \multicolumn{1}{c}{Difference} \\  
     \hline         
	 He &   -2.861154 &    -2.861153 &     0.000001 & &   -2.885608 &    -2.885608 &     0.000000  \\
	 Li  &  -7.432706 &    -7.432705 &     0.000001 & &   -7.443903 &    -7.443903 &     0.000000  \\
	 Be &  -14.572875 &   -14.572875 &     0.000000 & &  -14.598923 &   -14.598926 &    -0.000003  \\
	 B &  -24.532104 &   -24.532104 &     0.000000 &  & -24.566435 &   -24.566439 &    -0.000004  \\
	 C &  -37.691663 &   -37.691664 &    -0.000001 & &  -37.746778 &   -37.746781 &    -0.000003  \\
	 N &  -54.400883 &   -54.400885 &    -0.000002 & &  -54.482916 &   -54.482918 &    -0.000002  \\
	 O &  -74.811910 &   -74.811913 &    -0.000003 & &  -74.933839 &   -74.933844 &    -0.000005  \\
	 F &  -99.405657 &   -99.405660 &    -0.000003 & &  -99.576884 &   -99.576891 &    -0.000007  \\
	 Ne &  -128.532010 &  -128.532015 &    -0.000005 & & -128.760771 &  -128.760771 &     0.000000  \\
\hline\hline
\end{tabular}
\end{table*}

\section{Binding energy curves of H$_2$ and $\text{Ar}_2$ based on the PBE reference}
\label{app:pbe_ref}
\begin{figure}[h]
    \centering
    \begin{minipage}{0.46\textwidth}
        \includegraphics[width=1.\textwidth]{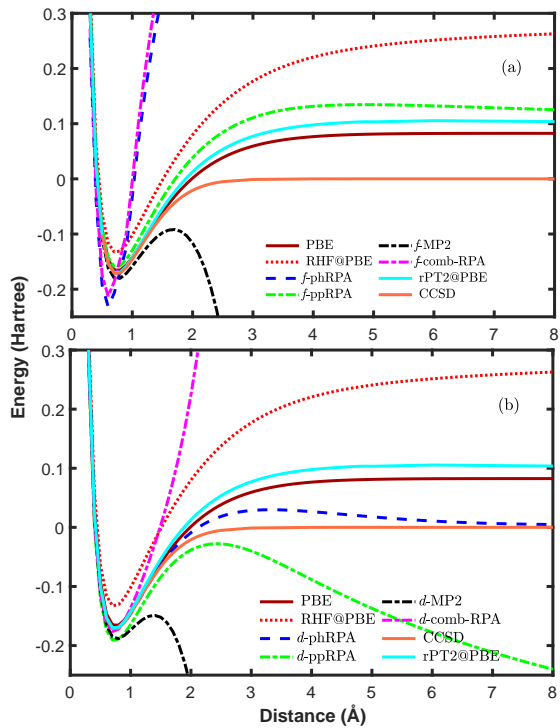} 
    \end{minipage}
        \caption{Binding energy curves for $\text{H}_{2}$ with the PBE reference and cc-pVTZ basis set. 
	     The plots here  differ from Fig.~\ref{fig:H2} only that, except for CCSD, all other methods are
	     based on the PBE reference instead of RHF.}
  \label{fig:H2_pbe}
\end{figure}
In Fig~\ref{fig:H2_pbe}, we present the binding energy curves for H$_2$ obtained with RPA, MP2, and rPT2 methods based on the PBE reference.

The CCSD result for H$_2$ is still the reference curve here. Comparing Fig.~\ref{fig:H2_pbe} to Fig.~\ref{fig:H2}, one can observe that the RPA methods on top of PBE 
in general yield less repulsive binding energy curves compared to their counterparts on top of RHF. 
The $f$-phRPA seems to be the exception in the sense that
$f$-phRPA@PBE curve is even more repulsive than $f$-phRPA@RHF. This unusual behavior also carries over to $f$-comb-RPA.
  
\begin{figure}[H]
    \centering
    \begin{minipage}{0.46\textwidth}
        \includegraphics[width=1.\textwidth]{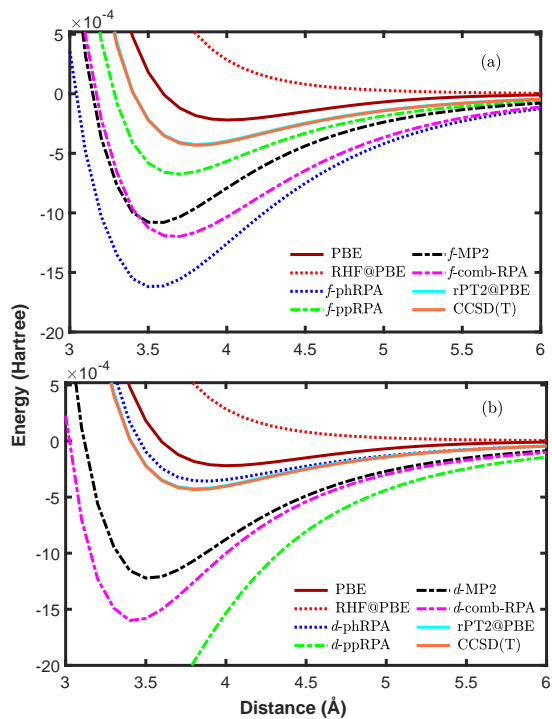} 
    \end{minipage}
        \caption{Binding energy curves for Ar$_2$ with the PBE reference and aug-cc-pVTZ basis set. 
	     The plots here  differ from Fig.~\ref{fig:Ar2} only that, except for CCSD(T), all other methods are
	     based on the PBE reference instead of RHF. The frozen-core approximation is used for the correlated methods.}
  \label{fig:Ar2_pbe}
\end{figure}

In Fig~\ref{fig:Ar2_pbe}, The binding energy curves for Ar$_2$ obtained with RPA, MP2, and rPT2 methods based on the PBE reference are presented.
Now the CCSD(T) curve is the reference curve to compare with. Compared to Fig.~\ref{fig:Ar2}, one can see that the RPA and MP2 curves show a pronounced dependence on
	the reference state, shifting downwards when moving from the the RHF reference to the PBE reference. In contrast with $f$-ppRPA@RHF which agrees 
	with the CCSD(T) result pretty well, now $f$-ppRPA@PBE overbinds the Ar$_2$ dimer substantially. It is even more so for $d$-ppRPA@PBE, with exchange
	contributions excluded. It is striking that the phRPA shows an opposite trend compared to the ppRPA, in that the $d$-phRPA@PBE is more repulsive 
	than $f$-phRPA@PBE. The comb-RPA curve now sits in between the ppRPA and phRPA curves, for both \textit{direct} and {full} flavors. The rPT2@PBE can 
	accurately reproduce the CCSD(T) curve, as already shown in Ref.~[\onlinecite{Ren/etal:2013}].

	From the RHF reference to the PBE reference, one can see that the RPA results have undergone substantial changes. A pertinent question is that 
	if an ``optimal" reference state can be found for practical RPA calculations. Recently, self-consistent phRPA schemes, in which an ``optimal"
	noninteracting reference is defined and iteratively optimized, are developed by Ye \textit{et al.} in the generalized optimized 
	effective potential framework \cite{Jin/etal:2017} and by Voora \textit{et al.} \cite{Voora/etal:2019} in the generalized KS framework. 
	It has been shown \cite{Jin/etal:2017,Voora/etal:2019} that the singles excitation effect \cite{Ren/etal:2011} can be automatically included in these schemes.

\end{appendix}

\bibliography{./CommonBib}

\begin{thebibliography}{86}
\expandafter\ifx\csname natexlab\endcsname\relax\def\natexlab#1{#1}\fi
\expandafter\ifx\csname bibnamefont\endcsname\relax
  \def\bibnamefont#1{#1}\fi
\expandafter\ifx\csname bibfnamefont\endcsname\relax
  \def\bibfnamefont#1{#1}\fi
\expandafter\ifx\csname citenamefont\endcsname\relax
  \def\citenamefont#1{#1}\fi
\expandafter\ifx\csname url\endcsname\relax
  \def\url#1{\texttt{#1}}\fi
\expandafter\ifx\csname urlprefix\endcsname\relax\def\urlprefix{URL }\fi
\providecommand{\bibinfo}[2]{#2}
\providecommand{\eprint}[2][]{\url{#2}}

\bibitem[{\citenamefont{Bohm and Pines}(1951)}]{Bohm/Pines:1951}
\bibinfo{author}{\bibfnamefont{D.}~\bibnamefont{Bohm}} \bibnamefont{and}
  \bibinfo{author}{\bibfnamefont{D.}~\bibnamefont{Pines}},
  \bibinfo{journal}{Phys. Rev.} \textbf{\bibinfo{volume}{82}},
  \bibinfo{pages}{625} (\bibinfo{year}{1951}).

\bibitem[{\citenamefont{Bohm and Pines}(1953)}]{Bohm/Pines:1953}
\bibinfo{author}{\bibfnamefont{D.}~\bibnamefont{Bohm}} \bibnamefont{and}
  \bibinfo{author}{\bibfnamefont{D.}~\bibnamefont{Pines}},
  \bibinfo{journal}{Phys. Rev.} \textbf{\bibinfo{volume}{92}},
  \bibinfo{pages}{609} (\bibinfo{year}{1953}).

\bibitem[{\citenamefont{He{\ss}elmann and
  G{\"o}rling}(2011{\natexlab{a}})}]{Hesselmann/Goerling:2011}
\bibinfo{author}{\bibfnamefont{A.}~\bibnamefont{He{\ss}elmann}}
  \bibnamefont{and}
  \bibinfo{author}{\bibfnamefont{A.}~\bibnamefont{G{\"o}rling}},
  \bibinfo{journal}{Mol. Phys.} \textbf{\bibinfo{volume}{109}},
  \bibinfo{pages}{2473} (\bibinfo{year}{2011}{\natexlab{a}}).

\bibitem[{\citenamefont{Eshuis et~al.}(2012)\citenamefont{Eshuis, Bates, and
  Furche}}]{Eshuis/Bates/Furche:2012}
\bibinfo{author}{\bibfnamefont{H.}~\bibnamefont{Eshuis}},
  \bibinfo{author}{\bibfnamefont{J.~E.} \bibnamefont{Bates}}, \bibnamefont{and}
  \bibinfo{author}{\bibfnamefont{F.}~\bibnamefont{Furche}},
  \bibinfo{journal}{Theor. Chem. Acc.} \textbf{\bibinfo{volume}{131}},
  \bibinfo{pages}{1084} (\bibinfo{year}{2012}).

\bibitem[{\citenamefont{Ren et~al.}(2012{\natexlab{a}})\citenamefont{Ren,
  Rinke, Joas, and Scheffler}}]{Ren/etal:2012b}
\bibinfo{author}{\bibfnamefont{X.}~\bibnamefont{Ren}},
  \bibinfo{author}{\bibfnamefont{P.}~\bibnamefont{Rinke}},
  \bibinfo{author}{\bibfnamefont{C.}~\bibnamefont{Joas}}, \bibnamefont{and}
  \bibinfo{author}{\bibfnamefont{M.}~\bibnamefont{Scheffler}},
  \bibinfo{journal}{J. Mater. Sci.} \textbf{\bibinfo{volume}{47}},
  \bibinfo{pages}{7447} (\bibinfo{year}{2012}{\natexlab{a}}).

\bibitem[{\citenamefont{Paier et~al.}(2012)\citenamefont{Paier, Ren, Rinke,
  Scuseria, Gr{\"u}neis, Kresse, and Scheffler}}]{Paier/etal:2012}
\bibinfo{author}{\bibfnamefont{J.}~\bibnamefont{Paier}},
  \bibinfo{author}{\bibfnamefont{X.}~\bibnamefont{Ren}},
  \bibinfo{author}{\bibfnamefont{P.}~\bibnamefont{Rinke}},
  \bibinfo{author}{\bibfnamefont{G.~E.} \bibnamefont{Scuseria}},
  \bibinfo{author}{\bibfnamefont{A.}~\bibnamefont{Gr{\"u}neis}},
  \bibinfo{author}{\bibfnamefont{G.}~\bibnamefont{Kresse}}, \bibnamefont{and}
  \bibinfo{author}{\bibfnamefont{M.}~\bibnamefont{Scheffler}},
  \bibinfo{journal}{New J. Phys.} \textbf{\bibinfo{volume}{14}},
  \bibinfo{pages}{043002} (\bibinfo{year}{2012}).

\bibitem[{\citenamefont{Furche}(2001)}]{Furche:2001}
\bibinfo{author}{\bibfnamefont{F.}~\bibnamefont{Furche}},
  \bibinfo{journal}{Phys. Rev. B} \textbf{\bibinfo{volume}{64}},
  \bibinfo{pages}{195120} (\bibinfo{year}{2001}).

\bibitem[{\citenamefont{Fuchs and Gonze}(2002)}]{Fuchs/Gonze:2002}
\bibinfo{author}{\bibfnamefont{M.}~\bibnamefont{Fuchs}} \bibnamefont{and}
  \bibinfo{author}{\bibfnamefont{X.}~\bibnamefont{Gonze}},
  \bibinfo{journal}{Phys. Rev. B} \textbf{\bibinfo{volume}{65}},
  \bibinfo{pages}{235109} (\bibinfo{year}{2002}).

\bibitem[{\citenamefont{Toulouse et~al.}(2009)\citenamefont{Toulouse, Gerber,
  Jansen, Savin, and \'Angy\'an}}]{Toulouse/etal:2009}
\bibinfo{author}{\bibfnamefont{J.}~\bibnamefont{Toulouse}},
  \bibinfo{author}{\bibfnamefont{I.~C.} \bibnamefont{Gerber}},
  \bibinfo{author}{\bibfnamefont{G.}~\bibnamefont{Jansen}},
  \bibinfo{author}{\bibfnamefont{A.}~\bibnamefont{Savin}}, \bibnamefont{and}
  \bibinfo{author}{\bibfnamefont{J.~G.} \bibnamefont{\'Angy\'an}},
  \bibinfo{journal}{Phys.\ Rev.\ Lett.} \textbf{\bibinfo{volume}{102}},
  \bibinfo{pages}{096404} (\bibinfo{year}{2009}).

\bibitem[{\citenamefont{Zhu et~al.}(2010)\citenamefont{Zhu, Toulouse, Savin,
  and {\'A}ngy{\'a}n}}]{Zhu/etal:2010}
\bibinfo{author}{\bibfnamefont{W.}~\bibnamefont{Zhu}},
  \bibinfo{author}{\bibfnamefont{J.}~\bibnamefont{Toulouse}},
  \bibinfo{author}{\bibfnamefont{A.}~\bibnamefont{Savin}}, \bibnamefont{and}
  \bibinfo{author}{\bibfnamefont{J.~G.} \bibnamefont{{\'A}ngy{\'a}n}},
  \bibinfo{journal}{J. Chem. Phys.} \textbf{\bibinfo{volume}{132}},
  \bibinfo{pages}{244108} (\bibinfo{year}{2010}).

\bibitem[{\citenamefont{He{\ss}elmann and
  G{\"o}rling}(2011{\natexlab{b}})}]{Hesselmann/Goerling:2011b}
\bibinfo{author}{\bibfnamefont{A.}~\bibnamefont{He{\ss}elmann}}
  \bibnamefont{and}
  \bibinfo{author}{\bibfnamefont{A.}~\bibnamefont{G{\"o}rling}},
  \bibinfo{journal}{Phys. Rev. Lett.} \textbf{\bibinfo{volume}{106}},
  \bibinfo{pages}{093001} (\bibinfo{year}{2011}{\natexlab{b}}).

\bibitem[{\citenamefont{Ren et~al.}(2011)\citenamefont{Ren, Tkatchenko, Rinke,
  and Scheffler}}]{Ren/etal:2011}
\bibinfo{author}{\bibfnamefont{X.}~\bibnamefont{Ren}},
  \bibinfo{author}{\bibfnamefont{A.}~\bibnamefont{Tkatchenko}},
  \bibinfo{author}{\bibfnamefont{P.}~\bibnamefont{Rinke}}, \bibnamefont{and}
  \bibinfo{author}{\bibfnamefont{M.}~\bibnamefont{Scheffler}},
  \bibinfo{journal}{Phys.\ Rev.\ Lett.} \textbf{\bibinfo{volume}{106}},
  \bibinfo{pages}{153003} (\bibinfo{year}{2011}).

\bibitem[{\citenamefont{Eshuis and Furche}(2011)}]{Eshuis/Furche:2011}
\bibinfo{author}{\bibfnamefont{H.}~\bibnamefont{Eshuis}} \bibnamefont{and}
  \bibinfo{author}{\bibfnamefont{F.}~\bibnamefont{Furche}},
  \bibinfo{journal}{J. Phys. Chem. Lett.} \textbf{\bibinfo{volume}{2}},
  \bibinfo{pages}{983} (\bibinfo{year}{2011}).

\bibitem[{\citenamefont{Harl and Kresse}(2008)}]{Harl/Kresse:2008}
\bibinfo{author}{\bibfnamefont{J.}~\bibnamefont{Harl}} \bibnamefont{and}
  \bibinfo{author}{\bibfnamefont{G.}~\bibnamefont{Kresse}},
  \bibinfo{journal}{Phys. Rev. B} \textbf{\bibinfo{volume}{77}},
  \bibinfo{pages}{045136} (\bibinfo{year}{2008}).

\bibitem[{\citenamefont{Harl and Kresse}(2009)}]{Harl/Kresse:2009}
\bibinfo{author}{\bibfnamefont{J.}~\bibnamefont{Harl}} \bibnamefont{and}
  \bibinfo{author}{\bibfnamefont{G.}~\bibnamefont{Kresse}},
  \bibinfo{journal}{Phys. Rev. Lett.} \textbf{\bibinfo{volume}{103}},
  \bibinfo{pages}{056401} (\bibinfo{year}{2009}).

\bibitem[{\citenamefont{Harl et~al.}(2010)\citenamefont{Harl, Schimka, and
  Kresse}}]{Harl/Schimka/Kresse:2010}
\bibinfo{author}{\bibfnamefont{J.}~\bibnamefont{Harl}},
  \bibinfo{author}{\bibfnamefont{L.}~\bibnamefont{Schimka}}, \bibnamefont{and}
  \bibinfo{author}{\bibfnamefont{G.}~\bibnamefont{Kresse}},
  \bibinfo{journal}{Phys. Rev. B} \textbf{\bibinfo{volume}{81}},
  \bibinfo{pages}{115126} (\bibinfo{year}{2010}).

\bibitem[{\citenamefont{Leb\`{e}gue et~al.}(2010)\citenamefont{Leb\`{e}gue,
  Harl, Gould, \'{A}ngy\'{a}n, Kresse, and Dobson}}]{Lebegue/etal:2010}
\bibinfo{author}{\bibfnamefont{S.}~\bibnamefont{Leb\`{e}gue}},
  \bibinfo{author}{\bibfnamefont{J.}~\bibnamefont{Harl}},
  \bibinfo{author}{\bibfnamefont{T.}~\bibnamefont{Gould}},
  \bibinfo{author}{\bibfnamefont{J.~G.} \bibnamefont{\'{A}ngy\'{a}n}},
  \bibinfo{author}{\bibfnamefont{G.}~\bibnamefont{Kresse}}, \bibnamefont{and}
  \bibinfo{author}{\bibfnamefont{J.~F.} \bibnamefont{Dobson}},
  \bibinfo{journal}{Phys. Rev. Lett.} \textbf{\bibinfo{volume}{105}},
  \bibinfo{pages}{196401} (\bibinfo{year}{2010}).

\bibitem[{\citenamefont{Nguyen and {de
  Gironcoli}}(2009)}]{Nguyen/deGironcoli:2009}
\bibinfo{author}{\bibfnamefont{H.-V.} \bibnamefont{Nguyen}} \bibnamefont{and}
  \bibinfo{author}{\bibfnamefont{S.}~\bibnamefont{{de Gironcoli}}},
  \bibinfo{journal}{Phys. Rev. B} \textbf{\bibinfo{volume}{79}},
  \bibinfo{pages}{205114} (\bibinfo{year}{2009}).

\bibitem[{\citenamefont{Lu et~al.}(2009)\citenamefont{Lu, Li, Rocca, and
  Galli}}]{Lu/Li/Rocca/Galli:2009}
\bibinfo{author}{\bibfnamefont{D.}~\bibnamefont{Lu}},
  \bibinfo{author}{\bibfnamefont{Y.}~\bibnamefont{Li}},
  \bibinfo{author}{\bibfnamefont{D.}~\bibnamefont{Rocca}}, \bibnamefont{and}
  \bibinfo{author}{\bibfnamefont{G.}~\bibnamefont{Galli}},
  \bibinfo{journal}{Phys. Rev. Lett.} \textbf{\bibinfo{volume}{102}},
  \bibinfo{pages}{206411} (\bibinfo{year}{2009}).

\bibitem[{\citenamefont{Casadei et~al.}({2012})\citenamefont{Casadei, Ren,
  Rinke, Rubio, and Scheffler}}]{Casadei/etal:2012}
\bibinfo{author}{\bibfnamefont{M.}~\bibnamefont{Casadei}},
  \bibinfo{author}{\bibfnamefont{X.}~\bibnamefont{Ren}},
  \bibinfo{author}{\bibfnamefont{P.}~\bibnamefont{Rinke}},
  \bibinfo{author}{\bibfnamefont{A.}~\bibnamefont{Rubio}}, \bibnamefont{and}
  \bibinfo{author}{\bibfnamefont{M.}~\bibnamefont{Scheffler}},
  \bibinfo{journal}{{Phys. Rev. Lett.}} \textbf{\bibinfo{volume}{{109}}},
  \bibinfo{pages}{146402} (\bibinfo{year}{{2012}}).

\bibitem[{\citenamefont{Casadei et~al.}(2016)\citenamefont{Casadei, Ren, Rinke,
  Rubio, and Scheffler}}]{Casadei/etal:2016}
\bibinfo{author}{\bibfnamefont{M.}~\bibnamefont{Casadei}},
  \bibinfo{author}{\bibfnamefont{X.}~\bibnamefont{Ren}},
  \bibinfo{author}{\bibfnamefont{P.}~\bibnamefont{Rinke}},
  \bibinfo{author}{\bibfnamefont{A.}~\bibnamefont{Rubio}}, \bibnamefont{and}
  \bibinfo{author}{\bibfnamefont{M.}~\bibnamefont{Scheffler}},
  \bibinfo{journal}{{Phys. Rev. B}} \textbf{\bibinfo{volume}{93}},
  \bibinfo{pages}{075153} (\bibinfo{year}{2016}).

\bibitem[{\citenamefont{Ren et~al.}(2009)\citenamefont{Ren, Rinke, and
  Scheffler}}]{Ren/etal:2009}
\bibinfo{author}{\bibfnamefont{X.}~\bibnamefont{Ren}},
  \bibinfo{author}{\bibfnamefont{P.}~\bibnamefont{Rinke}}, \bibnamefont{and}
  \bibinfo{author}{\bibfnamefont{M.}~\bibnamefont{Scheffler}},
  \bibinfo{journal}{Phys.\ Rev.\ B} \textbf{\bibinfo{volume}{80}},
  \bibinfo{pages}{045402} (\bibinfo{year}{2009}).

\bibitem[{\citenamefont{Schimka et~al.}(2010)\citenamefont{Schimka, Harl,
  Stroppa, Gr{\"u}neis, Marsman, Mittendorfer, and Kresse}}]{Schimka/etal:2010}
\bibinfo{author}{\bibfnamefont{L.}~\bibnamefont{Schimka}},
  \bibinfo{author}{\bibfnamefont{J.}~\bibnamefont{Harl}},
  \bibinfo{author}{\bibfnamefont{A.}~\bibnamefont{Stroppa}},
  \bibinfo{author}{\bibfnamefont{A.}~\bibnamefont{Gr{\"u}neis}},
  \bibinfo{author}{\bibfnamefont{M.}~\bibnamefont{Marsman}},
  \bibinfo{author}{\bibfnamefont{F.}~\bibnamefont{Mittendorfer}},
  \bibnamefont{and} \bibinfo{author}{\bibfnamefont{G.}~\bibnamefont{Kresse}},
  \bibinfo{journal}{Nature Materials} \textbf{\bibinfo{volume}{9}},
  \bibinfo{pages}{741} (\bibinfo{year}{2010}).

\bibitem[{\citenamefont{Mittendorfer et~al.}(2011)\citenamefont{Mittendorfer,
  Garhofer, Redinger, Klime\v{s}, Harl, and Kresse}}]{Mittendorfer/etal:2011}
\bibinfo{author}{\bibfnamefont{F.}~\bibnamefont{Mittendorfer}},
  \bibinfo{author}{\bibfnamefont{A.}~\bibnamefont{Garhofer}},
  \bibinfo{author}{\bibfnamefont{J.}~\bibnamefont{Redinger}},
  \bibinfo{author}{\bibfnamefont{J.}~\bibnamefont{Klime\v{s}}},
  \bibinfo{author}{\bibfnamefont{J.}~\bibnamefont{Harl}}, \bibnamefont{and}
  \bibinfo{author}{\bibfnamefont{G.}~\bibnamefont{Kresse}},
  \bibinfo{journal}{Phys Rev. B} \textbf{\bibinfo{volume}{84}},
  \bibinfo{pages}{201401(R)} (\bibinfo{year}{2011}).

\bibitem[{\citenamefont{Olsen et~al.}({2011})\citenamefont{Olsen, Yan,
  Mortensen, and Thygesen}}]{Olsen/etal:2011}
\bibinfo{author}{\bibfnamefont{T.}~\bibnamefont{Olsen}},
  \bibinfo{author}{\bibfnamefont{J.}~\bibnamefont{Yan}},
  \bibinfo{author}{\bibfnamefont{J.~J.} \bibnamefont{Mortensen}},
  \bibnamefont{and} \bibinfo{author}{\bibfnamefont{K.~S.}
  \bibnamefont{Thygesen}}, \bibinfo{journal}{{Phys. Rev. Lett.}}
  \textbf{\bibinfo{volume}{{107}}}, \bibinfo{pages}{{156401}}
  (\bibinfo{year}{{2011}}).

\bibitem[{\citenamefont{Bruneval}(2012)}]{Bruneval:2012}
\bibinfo{author}{\bibfnamefont{F.}~\bibnamefont{Bruneval}},
  \bibinfo{journal}{Phys. Rev. Lett.} \textbf{\bibinfo{volume}{108}},
  \bibinfo{pages}{256403} (\bibinfo{year}{2012}).

\bibitem[{\citenamefont{Kaltak et~al.}(2014{\natexlab{a}})\citenamefont{Kaltak,
  Klime\v{s}, and Kresse}}]{Kaltak/Klimes/Kresse:2014}
\bibinfo{author}{\bibfnamefont{M.}~\bibnamefont{Kaltak}},
  \bibinfo{author}{\bibfnamefont{J.}~\bibnamefont{Klime\v{s}}},
  \bibnamefont{and} \bibinfo{author}{\bibfnamefont{G.}~\bibnamefont{Kresse}},
  \bibinfo{journal}{J. Chem. Theory Comput.} \textbf{\bibinfo{volume}{10}},
  \bibinfo{pages}{2498} (\bibinfo{year}{2014}{\natexlab{a}}).

\bibitem[{\citenamefont{Kaltak et~al.}(2014{\natexlab{b}})\citenamefont{Kaltak,
  Klime\v{s}, and Kresse}}]{Kaltak/Klimes/Kresse:2014b}
\bibinfo{author}{\bibfnamefont{M.}~\bibnamefont{Kaltak}},
  \bibinfo{author}{\bibfnamefont{J.}~\bibnamefont{Klime\v{s}}},
  \bibnamefont{and} \bibinfo{author}{\bibfnamefont{G.}~\bibnamefont{Kresse}},
  \bibinfo{journal}{Phys. Rev. B} \textbf{\bibinfo{volume}{90}},
  \bibinfo{pages}{054115} (\bibinfo{year}{2014}{\natexlab{b}}).

\bibitem[{\citenamefont{Langreth and Perdew}(1975)}]{Langreth/Perdew:1975}
\bibinfo{author}{\bibfnamefont{D.~C.} \bibnamefont{Langreth}} \bibnamefont{and}
  \bibinfo{author}{\bibfnamefont{J.~P.} \bibnamefont{Perdew}},
  \bibinfo{journal}{Solid State Commun.} \textbf{\bibinfo{volume}{17}},
  \bibinfo{pages}{1425} (\bibinfo{year}{1975}).

\bibitem[{\citenamefont{Gunnarsson and
  Lundqvist}(1976)}]{Gunnarsson/Lundqvist:1976}
\bibinfo{author}{\bibfnamefont{O.}~\bibnamefont{Gunnarsson}} \bibnamefont{and}
  \bibinfo{author}{\bibfnamefont{B.~I.} \bibnamefont{Lundqvist}},
  \bibinfo{journal}{Phys.\ Rev.\ B} \textbf{\bibinfo{volume}{13}},
  \bibinfo{pages}{4274} (\bibinfo{year}{1976}).

\bibitem[{\citenamefont{Dobson}(1994)}]{Dobson:1994}
\bibinfo{author}{\bibfnamefont{J.~F.} \bibnamefont{Dobson}}, in
  \emph{\bibinfo{booktitle}{\textit{Topics in Condensed Matter Physics}}},
  edited by \bibinfo{editor}{\bibfnamefont{M.~P.} \bibnamefont{Das}}
  (\bibinfo{publisher}{Nova}, \bibinfo{address}{New York},
  \bibinfo{year}{1994}).

\bibitem[{\citenamefont{Scuseria et~al.}(2008)\citenamefont{Scuseria,
  Henderson, and Sorensen}}]{Scuseria/Henderson/Sorensen:2008}
\bibinfo{author}{\bibfnamefont{G.~E.} \bibnamefont{Scuseria}},
  \bibinfo{author}{\bibfnamefont{T.~M.} \bibnamefont{Henderson}},
  \bibnamefont{and} \bibinfo{author}{\bibfnamefont{D.~C.}
  \bibnamefont{Sorensen}}, \bibinfo{journal}{J. Chem. Phys.}
  \textbf{\bibinfo{volume}{129}}, \bibinfo{pages}{231101}
  (\bibinfo{year}{2008}).

\bibitem[{\citenamefont{Dahlen et~al.}(2005)\citenamefont{Dahlen, van Leeuwen,
  and von Barth}}]{Dahlen/Leeuwen/Barth:2005}
\bibinfo{author}{\bibfnamefont{N.~E.} \bibnamefont{Dahlen}},
  \bibinfo{author}{\bibfnamefont{R.}~\bibnamefont{van Leeuwen}},
  \bibnamefont{and} \bibinfo{author}{\bibfnamefont{U.}~\bibnamefont{von
  Barth}}, \bibinfo{journal}{Int. J. Quantum Chem.}
  \textbf{\bibinfo{volume}{101}}, \bibinfo{pages}{512} (\bibinfo{year}{2005}).

\bibitem[{\citenamefont{Ren et~al.}(2013)\citenamefont{Ren, Rinke, Scuseria,
  and Scheffler}}]{Ren/etal:2013}
\bibinfo{author}{\bibfnamefont{X.}~\bibnamefont{Ren}},
  \bibinfo{author}{\bibfnamefont{P.}~\bibnamefont{Rinke}},
  \bibinfo{author}{\bibfnamefont{G.~E.} \bibnamefont{Scuseria}},
  \bibnamefont{and}
  \bibinfo{author}{\bibfnamefont{M.}~\bibnamefont{Scheffler}},
  \bibinfo{journal}{Phys. Rev. B} \textbf{\bibinfo{volume}{88}},
  \bibinfo{pages}{035120} (\bibinfo{year}{2013}).

\bibitem[{\citenamefont{Bates and Furche}(2013)}]{Bates/Furche:2013}
\bibinfo{author}{\bibfnamefont{J.~E.} \bibnamefont{Bates}} \bibnamefont{and}
  \bibinfo{author}{\bibfnamefont{F.}~\bibnamefont{Furche}},
  \bibinfo{journal}{J. Chem. Phys.} \textbf{\bibinfo{volume}{139}},
  \bibinfo{pages}{171103} (\bibinfo{year}{2013}).

\bibitem[{\citenamefont{Chen et~al.}(2018)\citenamefont{Chen, Agee, and
  Furche}}]{Guo/Agee/Furche:2018}
\bibinfo{author}{\bibfnamefont{G.~P.} \bibnamefont{Chen}},
  \bibinfo{author}{\bibfnamefont{M.~M.} \bibnamefont{Agee}}, \bibnamefont{and}
  \bibinfo{author}{\bibfnamefont{F.}~\bibnamefont{Furche}},
  \bibinfo{journal}{J. Chem. Theo. Comput.} \textbf{\bibinfo{volume}{14}},
  \bibinfo{pages}{5701} (\bibinfo{year}{2018}).

\bibitem[{\citenamefont{Erhard et~al.}(2016)\citenamefont{Erhard, Bleiziffer,
  and G\"{o}rling}}]{Erhard/Bleiziffer/Goerling:2016}
\bibinfo{author}{\bibfnamefont{J.}~\bibnamefont{Erhard}},
  \bibinfo{author}{\bibfnamefont{P.}~\bibnamefont{Bleiziffer}},
  \bibnamefont{and}
  \bibinfo{author}{\bibfnamefont{A.}~\bibnamefont{G\"{o}rling}},
  \bibinfo{journal}{Phys. Rev. Lett.} \textbf{\bibinfo{volume}{117}},
  \bibinfo{pages}{143002} (\bibinfo{year}{2016}).

\bibitem[{\citenamefont{Ring and Schuck}(1980)}]{Ring/Schuck:1980}
\bibinfo{author}{\bibfnamefont{P.}~\bibnamefont{Ring}} \bibnamefont{and}
  \bibinfo{author}{\bibfnamefont{P.}~\bibnamefont{Schuck}},
  \emph{\bibinfo{title}{\textit{The Nuclear Many-Body Problem}}}
  (\bibinfo{publisher}{Springer}, \bibinfo{year}{1980}).

\bibitem[{\citenamefont{Blaizot and Ripka}(1986)}]{Blaizot/Ripka:1986}
\bibinfo{author}{\bibfnamefont{J.~P.} \bibnamefont{Blaizot}} \bibnamefont{and}
  \bibinfo{author}{\bibfnamefont{G.}~\bibnamefont{Ripka}},
  \emph{\bibinfo{title}{\textit{Quantum Theory of Finite Systems}}}
  (\bibinfo{publisher}{MIT Press}, \bibinfo{year}{1986}).

\bibitem[{\citenamefont{Gell-Mann and
  Brueckner}(1957)}]{Gell-Mann/Brueckner:1957}
\bibinfo{author}{\bibfnamefont{M.}~\bibnamefont{Gell-Mann}} \bibnamefont{and}
  \bibinfo{author}{\bibfnamefont{K.~A.} \bibnamefont{Brueckner}},
  \bibinfo{journal}{Phys. Rev.} \textbf{\bibinfo{volume}{106}},
  \bibinfo{pages}{364} (\bibinfo{year}{1957}).

\bibitem[{\citenamefont{{van Aggelen} et~al.}(2013)\citenamefont{{van Aggelen},
  Yang, and Yang}}]{Aggelen/etal:2013}
\bibinfo{author}{\bibfnamefont{H.}~\bibnamefont{{van Aggelen}}},
  \bibinfo{author}{\bibfnamefont{Y.}~\bibnamefont{Yang}}, \bibnamefont{and}
  \bibinfo{author}{\bibfnamefont{W.}~\bibnamefont{Yang}},
  \bibinfo{journal}{Phys. Rev. A} \textbf{\bibinfo{volume}{88}},
  \bibinfo{pages}{030501(R)} (\bibinfo{year}{2013}).

\bibitem[{\citenamefont{Peng et~al.}(2013)\citenamefont{Peng, Steinmann, {van
  Aggelen}, and Yang}}]{Peng/Steinmann/etal:2013}
\bibinfo{author}{\bibfnamefont{D.}~\bibnamefont{Peng}},
  \bibinfo{author}{\bibfnamefont{S.~N.} \bibnamefont{Steinmann}},
  \bibinfo{author}{\bibfnamefont{H.}~\bibnamefont{{van Aggelen}}},
  \bibnamefont{and} \bibinfo{author}{\bibfnamefont{W.}~\bibnamefont{Yang}},
  \bibinfo{journal}{J. Chem. Phys.} \textbf{\bibinfo{volume}{139}},
  \bibinfo{pages}{104112} (\bibinfo{year}{2013}).

\bibitem[{\citenamefont{Yang et~al.}(2013)\citenamefont{Yang, {van Aggelen},
  Steinmann, Peng, and Yang}}]{Yang/Aggelen/Steinmann/etal:2013}
\bibinfo{author}{\bibfnamefont{Y.}~\bibnamefont{Yang}},
  \bibinfo{author}{\bibfnamefont{H.}~\bibnamefont{{van Aggelen}}},
  \bibinfo{author}{\bibfnamefont{S.~N.} \bibnamefont{Steinmann}},
  \bibinfo{author}{\bibfnamefont{D.}~\bibnamefont{Peng}}, \bibnamefont{and}
  \bibinfo{author}{\bibfnamefont{W.}~\bibnamefont{Yang}}, \bibinfo{journal}{J.
  Chem. Phys.} \textbf{\bibinfo{volume}{139}}, \bibinfo{pages}{174110}
  (\bibinfo{year}{2013}).

\bibitem[{\citenamefont{{van Aggelen} et~al.}(2014)\citenamefont{{van Aggelen},
  Yang, and Yang}}]{Aggelen/etal:2014}
\bibinfo{author}{\bibfnamefont{H.}~\bibnamefont{{van Aggelen}}},
  \bibinfo{author}{\bibfnamefont{Y.}~\bibnamefont{Yang}}, \bibnamefont{and}
  \bibinfo{author}{\bibfnamefont{W.}~\bibnamefont{Yang}}, \bibinfo{journal}{J.
  Chem. Phys.} \textbf{\bibinfo{volume}{140}}, \bibinfo{pages}{18A511}
  (\bibinfo{year}{2014}).

\bibitem[{\citenamefont{Peng et~al.}(2014)\citenamefont{Peng, van Aggelen,
  Yang, and Yang}}]{Peng/etal:2014}
\bibinfo{author}{\bibfnamefont{D.}~\bibnamefont{Peng}},
  \bibinfo{author}{\bibfnamefont{H.}~\bibnamefont{van Aggelen}},
  \bibinfo{author}{\bibfnamefont{Y.}~\bibnamefont{Yang}}, \bibnamefont{and}
  \bibinfo{author}{\bibfnamefont{W.}~\bibnamefont{Yang}}, \bibinfo{journal}{J.
  Chem. Phys.} \textbf{\bibinfo{volume}{140}}, \bibinfo{pages}{18A522}
  (\bibinfo{year}{2014}).

\bibitem[{\citenamefont{Scuseria et~al.}(2013)\citenamefont{Scuseria,
  Henderson, and Bulik}}]{Scuseria/Henderson/Bulik:2013}
\bibinfo{author}{\bibfnamefont{E.}~\bibnamefont{Scuseria}},
  \bibinfo{author}{\bibfnamefont{T.~M.} \bibnamefont{Henderson}},
  \bibnamefont{and} \bibinfo{author}{\bibfnamefont{I.~W.} \bibnamefont{Bulik}},
  \bibinfo{journal}{J. Chem. Phys.} \textbf{\bibinfo{volume}{139}},
  \bibinfo{pages}{104113} (\bibinfo{year}{2013}).

\bibitem[{\citenamefont{Luttinger and Ward}(1960)}]{Luttinger/Ward:1961}
\bibinfo{author}{\bibfnamefont{J.~M.} \bibnamefont{Luttinger}}
  \bibnamefont{and} \bibinfo{author}{\bibfnamefont{J.~C.} \bibnamefont{Ward}},
  \bibinfo{journal}{Phys. Rev.} \textbf{\bibinfo{volume}{118}},
  \bibinfo{pages}{1417} (\bibinfo{year}{1960}).

\bibitem[{\citenamefont{Hedin}(1965)}]{Hedin:1965}
\bibinfo{author}{\bibfnamefont{L.}~\bibnamefont{Hedin}},
  \bibinfo{journal}{Phys.\ Rev.} \textbf{\bibinfo{volume}{139}},
  \bibinfo{pages}{A796} (\bibinfo{year}{1965}).

\bibitem[{\citenamefont{Baym and Kadanoff}(1961)}]{Baym/Kadanoff:1961}
\bibinfo{author}{\bibfnamefont{G.}~\bibnamefont{Baym}} \bibnamefont{and}
  \bibinfo{author}{\bibfnamefont{L.~P.} \bibnamefont{Kadanoff}},
  \bibinfo{journal}{Phys. Rev.} \textbf{\bibinfo{volume}{124}},
  \bibinfo{pages}{287} (\bibinfo{year}{1961}).

\bibitem[{\citenamefont{Kanamori}(1963)}]{Kanamori:1963}
\bibinfo{author}{\bibfnamefont{J.}~\bibnamefont{Kanamori}},
  \bibinfo{journal}{Prog. Theor. Phys.} \textbf{\bibinfo{volume}{30}},
  \bibinfo{pages}{275} (\bibinfo{year}{1963}).

\bibitem[{\citenamefont{Romaniello et~al.}(2012)\citenamefont{Romaniello,
  Bechstedt, and Reining}}]{Romaniello/Bechstedt/Reining:2012}
\bibinfo{author}{\bibfnamefont{P.}~\bibnamefont{Romaniello}},
  \bibinfo{author}{\bibfnamefont{F.}~\bibnamefont{Bechstedt}},
  \bibnamefont{and} \bibinfo{author}{\bibfnamefont{L.}~\bibnamefont{Reining}},
  \bibinfo{journal}{Phys. Rev. B} \textbf{\bibinfo{volume}{85}},
  \bibinfo{pages}{155131} (\bibinfo{year}{2012}).

\bibitem[{\citenamefont{Zhang et~al.}(2017)\citenamefont{Zhang, Su, and
  Yang}}]{Zhang/Su/Yang:2017}
\bibinfo{author}{\bibfnamefont{D.}~\bibnamefont{Zhang}},
  \bibinfo{author}{\bibfnamefont{N.~Q.} \bibnamefont{Su}}, \bibnamefont{and}
  \bibinfo{author}{\bibfnamefont{W.}~\bibnamefont{Yang}}, \bibinfo{journal}{J.
  Phys. Chem. Lett.} \textbf{\bibinfo{volume}{8}}, \bibinfo{pages}{3223}
  (\bibinfo{year}{2017}).

\bibitem[{\citenamefont{Bethe and Goldstone}(1957)}]{Bethe/Goldstone:1957}
\bibinfo{author}{\bibfnamefont{H.~A.} \bibnamefont{Bethe}} \bibnamefont{and}
  \bibinfo{author}{\bibfnamefont{J.}~\bibnamefont{Goldstone}},
  \bibinfo{journal}{Proc. R. Soc. London A} \textbf{\bibinfo{volume}{238}},
  \bibinfo{pages}{551} (\bibinfo{year}{1957}).

\bibitem[{\citenamefont{Blum et~al.}(2009)\citenamefont{Blum, Hanke, Gehrke,
  Havu, Havu, Ren, Reuter, and Scheffler}}]{Blum/etal:2009}
\bibinfo{author}{\bibfnamefont{V.}~\bibnamefont{Blum}},
  \bibinfo{author}{\bibfnamefont{F.}~\bibnamefont{Hanke}},
  \bibinfo{author}{\bibfnamefont{R.}~\bibnamefont{Gehrke}},
  \bibinfo{author}{\bibfnamefont{P.}~\bibnamefont{Havu}},
  \bibinfo{author}{\bibfnamefont{V.}~\bibnamefont{Havu}},
  \bibinfo{author}{\bibfnamefont{X.}~\bibnamefont{Ren}},
  \bibinfo{author}{\bibfnamefont{K.}~\bibnamefont{Reuter}}, \bibnamefont{and}
  \bibinfo{author}{\bibfnamefont{M.}~\bibnamefont{Scheffler}},
  \bibinfo{journal}{Comp. Phys. Comm.} \textbf{\bibinfo{volume}{180}},
  \bibinfo{pages}{2175} (\bibinfo{year}{2009}).

\bibitem[{\citenamefont{Ren et~al.}(2012{\natexlab{b}})\citenamefont{Ren,
  Rinke, Blum, Wieferink, Tkatchenko, Sanfilippo, Reuter, and
  Scheffler}}]{Ren/etal:2012}
\bibinfo{author}{\bibfnamefont{X.}~\bibnamefont{Ren}},
  \bibinfo{author}{\bibfnamefont{P.}~\bibnamefont{Rinke}},
  \bibinfo{author}{\bibfnamefont{V.}~\bibnamefont{Blum}},
  \bibinfo{author}{\bibfnamefont{J.}~\bibnamefont{Wieferink}},
  \bibinfo{author}{\bibfnamefont{A.}~\bibnamefont{Tkatchenko}},
  \bibinfo{author}{\bibfnamefont{A.}~\bibnamefont{Sanfilippo}},
  \bibinfo{author}{\bibfnamefont{K.}~\bibnamefont{Reuter}}, \bibnamefont{and}
  \bibinfo{author}{\bibfnamefont{M.}~\bibnamefont{Scheffler}},
  \bibinfo{journal}{New J. Phys.} \textbf{\bibinfo{volume}{14}},
  \bibinfo{pages}{053020} (\bibinfo{year}{2012}{\natexlab{b}}).

\bibitem[{\citenamefont{McLachlan and Ball}(1964)}]{McLachlan/Ball:1964}
\bibinfo{author}{\bibfnamefont{A.~D.} \bibnamefont{McLachlan}}
  \bibnamefont{and} \bibinfo{author}{\bibfnamefont{M.~A.} \bibnamefont{Ball}},
  \bibinfo{journal}{Rev. Mol. Phys.} \textbf{\bibinfo{volume}{36}},
  \bibinfo{pages}{844} (\bibinfo{year}{1964}).

\bibitem[{\citenamefont{Oddershede et~al.}(1975)\citenamefont{Oddershede,
  J\"{o}rgensen, and Beebe}}]{Oddershede/etal:1975}
\bibinfo{author}{\bibfnamefont{J.}~\bibnamefont{Oddershede}},
  \bibinfo{author}{\bibfnamefont{P.}~\bibnamefont{J\"{o}rgensen}},
  \bibnamefont{and} \bibinfo{author}{\bibfnamefont{N.~H.~F.}
  \bibnamefont{Beebe}}, \bibinfo{journal}{J. Chem. Phys.}
  \textbf{\bibinfo{volume}{63}}, \bibinfo{pages}{2996} (\bibinfo{year}{1975}).

\bibitem[{\citenamefont{{\'A}ngy{\'a}n
  et~al.}(2011)\citenamefont{{\'A}ngy{\'a}n, Liu, Toulouse, and
  Jansen}}]{Angyan/etal:2011}
\bibinfo{author}{\bibfnamefont{J.~G.} \bibnamefont{{\'A}ngy{\'a}n}},
  \bibinfo{author}{\bibfnamefont{R.-F.} \bibnamefont{Liu}},
  \bibinfo{author}{\bibfnamefont{J.}~\bibnamefont{Toulouse}}, \bibnamefont{and}
  \bibinfo{author}{\bibfnamefont{G.}~\bibnamefont{Jansen}},
  \bibinfo{journal}{J. Chem. Theory Comput.} \textbf{\bibinfo{volume}{7}},
  \bibinfo{pages}{3116} (\bibinfo{year}{2011}).

\bibitem[{\citenamefont{Seeger and Pople}(1977)}]{Rolf/Seeger:1977}
\bibinfo{author}{\bibfnamefont{R.}~\bibnamefont{Seeger}} \bibnamefont{and}
  \bibinfo{author}{\bibfnamefont{J.~A.} \bibnamefont{Pople}},
  \bibinfo{journal}{J.Chem. Phys.} \textbf{\bibinfo{volume}{66}},
  \bibinfo{pages}{3045} (\bibinfo{year}{1977}).

\bibitem[{\citenamefont{Eshuis et~al.}(2010)\citenamefont{Eshuis, Yarkony, and
  Furche}}]{Eshuis/Yarkony/Furche:2010}
\bibinfo{author}{\bibfnamefont{H.}~\bibnamefont{Eshuis}},
  \bibinfo{author}{\bibfnamefont{J.}~\bibnamefont{Yarkony}}, \bibnamefont{and}
  \bibinfo{author}{\bibfnamefont{F.}~\bibnamefont{Furche}},
  \bibinfo{journal}{J. Chem. Phys.} \textbf{\bibinfo{volume}{132}},
  \bibinfo{pages}{234114} (\bibinfo{year}{2010}).

\bibitem[{\citenamefont{Mori-Sanchez et~al.}(2012)\citenamefont{Mori-Sanchez,
  Cohen, and Yang}}]{Mori-Sanchez/etal:2012}
\bibinfo{author}{\bibfnamefont{P.}~\bibnamefont{Mori-Sanchez}},
  \bibinfo{author}{\bibfnamefont{A.~J.} \bibnamefont{Cohen}}, \bibnamefont{and}
  \bibinfo{author}{\bibfnamefont{W.}~\bibnamefont{Yang}},
  \bibinfo{journal}{Phys. Rev. A} \textbf{\bibinfo{volume}{85}},
  \bibinfo{pages}{042507} (\bibinfo{year}{2012}).

\bibitem[{\citenamefont{He{\ss}elmann}(2011)}]{Hesselmann:2011}
\bibinfo{author}{\bibfnamefont{A.}~\bibnamefont{He{\ss}elmann}},
  \bibinfo{journal}{J. Chem. Phys.} \textbf{\bibinfo{volume}{134}},
  \bibinfo{pages}{204107} (\bibinfo{year}{2011}).

\bibitem[{\citenamefont{Furche}(2008)}]{Furche:2008}
\bibinfo{author}{\bibfnamefont{F.}~\bibnamefont{Furche}}, \bibinfo{journal}{J.
  Chem. Phys.} \textbf{\bibinfo{volume}{129}}, \bibinfo{pages}{114105}
  (\bibinfo{year}{2008}).

\bibitem[{\citenamefont{Szabo and Ostlund}(1989)}]{Szabo/Ostlund:1989}
\bibinfo{author}{\bibfnamefont{A.}~\bibnamefont{Szabo}} \bibnamefont{and}
  \bibinfo{author}{\bibfnamefont{N.~S.} \bibnamefont{Ostlund}},
  \emph{\bibinfo{title}{\textit{Modern Quantum Chemistry: Introduction to
  Advanced Electronic Structure Theory}}} (\bibinfo{publisher}{McGraw-Hill},
  \bibinfo{address}{New York}, \bibinfo{year}{1989}).

\bibitem[{\citenamefont{M{\o}ller and Plesset}(1934)}]{Moller/Plesset:1934}
\bibinfo{author}{\bibfnamefont{C.}~\bibnamefont{M{\o}ller}} \bibnamefont{and}
  \bibinfo{author}{\bibfnamefont{M.~S.} \bibnamefont{Plesset}},
  \bibinfo{journal}{Phys. Rev.} \textbf{\bibinfo{volume}{46}},
  \bibinfo{pages}{618} (\bibinfo{year}{1934}).

\bibitem[{\citenamefont{C{\'i}zek}(1966)}]{Cizek:1966}
\bibinfo{author}{\bibfnamefont{J.}~\bibnamefont{C{\'i}zek}},
  \bibinfo{journal}{J. Chem. Phys.} \textbf{\bibinfo{volume}{45}},
  \bibinfo{pages}{4256} (\bibinfo{year}{1966}).

\bibitem[{\citenamefont{Bishop and L{\"u}hrmann}(1978)}]{Bishop/Luehrmann:1978}
\bibinfo{author}{\bibfnamefont{R.~F.} \bibnamefont{Bishop}} \bibnamefont{and}
  \bibinfo{author}{\bibfnamefont{K.~H.} \bibnamefont{L{\"u}hrmann}},
  \bibinfo{journal}{Phys. Rev. B} \textbf{\bibinfo{volume}{17}},
  \bibinfo{pages}{3757} (\bibinfo{year}{1978}).

\bibitem[{\citenamefont{Shepherd et~al.}(2014)\citenamefont{Shepherd,
  Henderson, and Scuseria}}]{Shepherd/Henderson/Scuseria/:2014}
\bibinfo{author}{\bibfnamefont{J.~J.} \bibnamefont{Shepherd}},
  \bibinfo{author}{\bibfnamefont{T.~M.} \bibnamefont{Henderson}},
  \bibnamefont{and} \bibinfo{author}{\bibfnamefont{G.~E.}
  \bibnamefont{Scuseria}}, \bibinfo{journal}{Phys. Rev. Lett.}
  \textbf{\bibinfo{volume}{112}}, \bibinfo{pages}{133002}
  (\bibinfo{year}{2014}).

\bibitem[{\citenamefont{Havu et~al.}(2009)\citenamefont{Havu, Blum, Havu, and
  Scheffler}}]{Havu/etal:2009}
\bibinfo{author}{\bibfnamefont{V.}~\bibnamefont{Havu}},
  \bibinfo{author}{\bibfnamefont{V.}~\bibnamefont{Blum}},
  \bibinfo{author}{\bibfnamefont{P.}~\bibnamefont{Havu}}, \bibnamefont{and}
  \bibinfo{author}{\bibfnamefont{M.}~\bibnamefont{Scheffler}},
  \bibinfo{journal}{J. Comp. Phys.} \textbf{\bibinfo{volume}{228}},
  \bibinfo{pages}{8367} (\bibinfo{year}{2009}).

\bibitem[{\citenamefont{Bartlett and Musia{\l}}(2007)}]{Bartlett/Musial:2007}
\bibinfo{author}{\bibfnamefont{R.~J.} \bibnamefont{Bartlett}} \bibnamefont{and}
  \bibinfo{author}{\bibfnamefont{M.}~\bibnamefont{Musia{\l}}},
  \bibinfo{journal}{Rev. Mod. Phys.} \textbf{\bibinfo{volume}{79}},
  \bibinfo{pages}{291} (\bibinfo{year}{2007}).

\bibitem[{\citenamefont{Werner and Knowles}(1988)}]{Werner/Knowles:1988}
\bibinfo{author}{\bibfnamefont{H.~J.} \bibnamefont{Werner}} \bibnamefont{and}
  \bibinfo{author}{\bibfnamefont{P.~J.} \bibnamefont{Knowles}},
  \bibinfo{journal}{J. Chem. Phys.} \textbf{\bibinfo{volume}{89}},
  \bibinfo{pages}{5803} (\bibinfo{year}{1988}).

\bibitem[{\citenamefont{Frisch et~al.}()\citenamefont{Frisch, Trucks, Schlegel,
  Scuseria, Robb, Cheeseman, Scalmani, Barone, Mennucci, Petersson
  et~al.}}]{g09}
\bibinfo{author}{\bibfnamefont{M.~J.} \bibnamefont{Frisch}},
  \bibinfo{author}{\bibfnamefont{G.~W.} \bibnamefont{Trucks}},
  \bibinfo{author}{\bibfnamefont{H.~B.} \bibnamefont{Schlegel}},
  \bibinfo{author}{\bibfnamefont{G.~E.} \bibnamefont{Scuseria}},
  \bibinfo{author}{\bibfnamefont{M.~A.} \bibnamefont{Robb}},
  \bibinfo{author}{\bibfnamefont{J.~R.} \bibnamefont{Cheeseman}},
  \bibinfo{author}{\bibfnamefont{G.}~\bibnamefont{Scalmani}},
  \bibinfo{author}{\bibfnamefont{V.}~\bibnamefont{Barone}},
  \bibinfo{author}{\bibfnamefont{B.}~\bibnamefont{Mennucci}},
  \bibinfo{author}{\bibfnamefont{G.~A.} \bibnamefont{Petersson}},
  \bibnamefont{et~al.}, \emph{\bibinfo{title}{Gaussian∼09{W} {R}evision
  {D}.01}}, \bibinfo{note}{gaussian Inc. Wallingford CT 2009}.

\bibitem[{\citenamefont{{Shen} et~al.}(2018)\citenamefont{{Shen}, {Zhang}, and
  {Scheffler}}}]{2018arXiv181008142S}
\bibinfo{author}{\bibfnamefont{T.}~\bibnamefont{{Shen}}},
  \bibinfo{author}{\bibfnamefont{I.~Y.} \bibnamefont{{Zhang}}},
  \bibnamefont{and}
  \bibinfo{author}{\bibfnamefont{M.}~\bibnamefont{{Scheffler}}},
  \bibinfo{journal}{arXiv e-prints} \bibinfo{eid}{arXiv:1810.08142}
  (\bibinfo{year}{2018}), \eprint{1810.08142}.

\bibitem[{\citenamefont{Lischka et~al.}(2017)\citenamefont{Lischka, Shepard,
  Shavitt, Pitzer, Dallos, M{\"u}ller, Szalay, Brown, Ahlrichs, Böhm
  et~al.}}]{COLUMBUS}
\bibinfo{author}{\bibfnamefont{H.}~\bibnamefont{Lischka}},
  \bibinfo{author}{\bibfnamefont{R.}~\bibnamefont{Shepard}},
  \bibinfo{author}{\bibfnamefont{I.}~\bibnamefont{Shavitt}},
  \bibinfo{author}{\bibfnamefont{R.~M.} \bibnamefont{Pitzer}},
  \bibinfo{author}{\bibfnamefont{M.}~\bibnamefont{Dallos}},
  \bibinfo{author}{\bibfnamefont{T.}~\bibnamefont{M{\"u}ller}},
  \bibinfo{author}{\bibfnamefont{P.~G.} \bibnamefont{Szalay}},
  \bibinfo{author}{\bibfnamefont{F.~B.} \bibnamefont{Brown}},
  \bibinfo{author}{\bibfnamefont{R.}~\bibnamefont{Ahlrichs}},
  \bibinfo{author}{\bibfnamefont{H.~J.} \bibnamefont{Böhm}},
  \bibnamefont{et~al.}, \emph{\bibinfo{title}{Columbus {R}elease~7.0}}
  (\bibinfo{year}{2017}), \bibinfo{note}{an ab initio electronic structure
  program}.

\bibitem[{\citenamefont{T.~H.~Dunning}(1989)}]{Dunning:1989}
\bibinfo{author}{\bibfnamefont{J.}~\bibnamefont{T.~H.~Dunning}},
  \bibinfo{journal}{J. Chem. Phys.} \textbf{\bibinfo{volume}{90}},
  \bibinfo{pages}{1007} (\bibinfo{year}{1989}).

\bibitem[{\citenamefont{Perdew et~al.}(1996)\citenamefont{Perdew, Burke, and
  Ernzerhof}}]{Perdew/Burke/Ernzerhof:1996}
\bibinfo{author}{\bibfnamefont{J.~P.} \bibnamefont{Perdew}},
  \bibinfo{author}{\bibfnamefont{K.}~\bibnamefont{Burke}}, \bibnamefont{and}
  \bibinfo{author}{\bibfnamefont{M.}~\bibnamefont{Ernzerhof}},
  \bibinfo{journal}{Phys. Rev. Lett} \textbf{\bibinfo{volume}{77}},
  \bibinfo{pages}{3865} (\bibinfo{year}{1996}).

\bibitem[{\citenamefont{Cohen et~al.}(2012)\citenamefont{Cohen, Mori-S\'anchez,
  and Yang}}]{Cohen/etal:2012}
\bibinfo{author}{\bibfnamefont{A.~J.} \bibnamefont{Cohen}},
  \bibinfo{author}{\bibfnamefont{P.}~\bibnamefont{Mori-S\'anchez}},
  \bibnamefont{and} \bibinfo{author}{\bibfnamefont{W.}~\bibnamefont{Yang}},
  \bibinfo{journal}{Chem. Rev.} \textbf{\bibinfo{volume}{112}},
  \bibinfo{pages}{289} (\bibinfo{year}{2012}).

\bibitem[{\citenamefont{Zhang et~al.}(2016{\natexlab{a}})\citenamefont{Zhang,
  Rinke, Perdew, and Scheffler}}]{IgorZhang/etal:2016a}
\bibinfo{author}{\bibfnamefont{I.~Y.} \bibnamefont{Zhang}},
  \bibinfo{author}{\bibfnamefont{P.}~\bibnamefont{Rinke}},
  \bibinfo{author}{\bibfnamefont{J.~P.} \bibnamefont{Perdew}},
  \bibnamefont{and}
  \bibinfo{author}{\bibfnamefont{M.}~\bibnamefont{Scheffler}},
  \bibinfo{journal}{Phys. Rev. Lett.} \textbf{\bibinfo{volume}{117}},
  \bibinfo{pages}{133002} (\bibinfo{year}{2016}{\natexlab{a}}).

\bibitem[{\citenamefont{Zhang et~al.}(2016{\natexlab{b}})\citenamefont{Zhang,
  Rinke, and Scheffler}}]{IgorZhang/etal:2016b}
\bibinfo{author}{\bibfnamefont{I.~Y.} \bibnamefont{Zhang}},
  \bibinfo{author}{\bibfnamefont{P.}~\bibnamefont{Rinke}}, \bibnamefont{and}
  \bibinfo{author}{\bibfnamefont{M.}~\bibnamefont{Scheffler}},
  \bibinfo{journal}{New J. Phys.} \textbf{\bibinfo{volume}{18}},
  \bibinfo{pages}{073026} (\bibinfo{year}{2016}{\natexlab{b}}).

\bibitem[{\citenamefont{Gr{\"u}neis et~al.}(2009)\citenamefont{Gr{\"u}neis,
  Marsman, Harl, Schimka, and Kresse}}]{Grueneis/etal:2009}
\bibinfo{author}{\bibfnamefont{A.}~\bibnamefont{Gr{\"u}neis}},
  \bibinfo{author}{\bibfnamefont{M.}~\bibnamefont{Marsman}},
  \bibinfo{author}{\bibfnamefont{J.}~\bibnamefont{Harl}},
  \bibinfo{author}{\bibfnamefont{L.}~\bibnamefont{Schimka}}, \bibnamefont{and}
  \bibinfo{author}{\bibfnamefont{G.}~\bibnamefont{Kresse}},
  \bibinfo{journal}{J. Chem. Phys.} \textbf{\bibinfo{volume}{131}},
  \bibinfo{pages}{154115} (\bibinfo{year}{2009}).

\bibitem[{\citenamefont{Paier et~al.}(2010)\citenamefont{Paier, Janesko,
  Henderson, Scuseria, Gr{\"u}neis, and Kresse}}]{Paier/etal:2010}
\bibinfo{author}{\bibfnamefont{J.}~\bibnamefont{Paier}},
  \bibinfo{author}{\bibfnamefont{B.~G.} \bibnamefont{Janesko}},
  \bibinfo{author}{\bibfnamefont{T.~M.} \bibnamefont{Henderson}},
  \bibinfo{author}{\bibfnamefont{G.~E.} \bibnamefont{Scuseria}},
  \bibinfo{author}{\bibfnamefont{A.}~\bibnamefont{Gr{\"u}neis}},
  \bibnamefont{and} \bibinfo{author}{\bibfnamefont{G.}~\bibnamefont{Kresse}},
  \bibinfo{journal}{J. Chem. Phys.} \textbf{\bibinfo{volume}{132}},
  \bibinfo{pages}{094103} (\bibinfo{year}{2010}), \bibinfo{note}{erratum: {\it
  ibid.}~{\bf 133}, 179902 (2010).}

\bibitem[{\citenamefont{Coulson and Fischer}(1949)}]{Coulson/Fischer:1949}
\bibinfo{author}{\bibfnamefont{C.}~\bibnamefont{Coulson}} \bibnamefont{and}
  \bibinfo{author}{\bibfnamefont{I.}~\bibnamefont{Fischer}},
  \bibinfo{journal}{Phil. Mag.} \textbf{\bibinfo{volume}{40}},
  \bibinfo{pages}{386} (\bibinfo{year}{1949}).

\bibitem[{\citenamefont{Henderson and
  Scuseria}(2010)}]{Henderson/Scuseria:2010}
\bibinfo{author}{\bibfnamefont{T.~M.} \bibnamefont{Henderson}}
  \bibnamefont{and} \bibinfo{author}{\bibfnamefont{G.~E.}
  \bibnamefont{Scuseria}}, \bibinfo{journal}{Mol. Phys.}
  \textbf{\bibinfo{volume}{108}}, \bibinfo{pages}{2511} (\bibinfo{year}{2010}).

\bibitem[{\citenamefont{Ihrig et~al.}(2015)\citenamefont{Ihrig, Wieferink,
  Zhang, Ropo, Ren, Rinke, Scheffler, and Blum}}]{Ihrig/etal:2015}
\bibinfo{author}{\bibfnamefont{A.~C.} \bibnamefont{Ihrig}},
  \bibinfo{author}{\bibfnamefont{J.}~\bibnamefont{Wieferink}},
  \bibinfo{author}{\bibfnamefont{I.~Y.} \bibnamefont{Zhang}},
  \bibinfo{author}{\bibfnamefont{M.}~\bibnamefont{Ropo}},
  \bibinfo{author}{\bibfnamefont{X.}~\bibnamefont{Ren}},
  \bibinfo{author}{\bibfnamefont{P.}~\bibnamefont{Rinke}},
  \bibinfo{author}{\bibfnamefont{M.}~\bibnamefont{Scheffler}},
  \bibnamefont{and} \bibinfo{author}{\bibfnamefont{V.}~\bibnamefont{Blum}},
  \bibinfo{journal}{New J. Phys.} \textbf{\bibinfo{volume}{17}},
  \bibinfo{pages}{093020} (\bibinfo{year}{2015}).

\bibitem[{\citenamefont{Jin et~al.}(2017)\citenamefont{Jin, Zhang, Chen, Su,
  and Yang}}]{Jin/etal:2017}
\bibinfo{author}{\bibfnamefont{Y.}~\bibnamefont{Jin}},
  \bibinfo{author}{\bibfnamefont{D.}~\bibnamefont{Zhang}},
  \bibinfo{author}{\bibfnamefont{Z.}~\bibnamefont{Chen}},
  \bibinfo{author}{\bibfnamefont{N.~Q.} \bibnamefont{Su}}, \bibnamefont{and}
  \bibinfo{author}{\bibfnamefont{W.}~\bibnamefont{Yang}}, \bibinfo{journal}{J.
  Phys. Chem. Lett.} \textbf{\bibinfo{volume}{8}}, \bibinfo{pages}{4746}
  (\bibinfo{year}{2017}).

\bibitem[{\citenamefont{Voora et~al.}(2019)\citenamefont{Voora, Balasubramani,
  and Furche}}]{Voora/etal:2019}
\bibinfo{author}{\bibfnamefont{V.~K.} \bibnamefont{Voora}},
  \bibinfo{author}{\bibfnamefont{S.~G.} \bibnamefont{Balasubramani}},
  \bibnamefont{and} \bibinfo{author}{\bibfnamefont{F.}~\bibnamefont{Furche}},
  \bibinfo{journal}{Phys. Rev. A} \textbf{\bibinfo{volume}{99}},
  \bibinfo{pages}{012518} (\bibinfo{year}{2019}).

\end{thebibliography}

\end{document}